\documentclass[11pt,fullpage]{article}

\usepackage{amsmath}
\usepackage{amssymb}
\usepackage{amsfonts}
\usepackage{algorithmic}
\usepackage{algorithm}
\usepackage{graphicx}
\usepackage{subfig}
\usepackage{paralist}
\usepackage{color}
\usepackage{multirow}
\usepackage{enumitem}
\usepackage{comment}
\usepackage{balance}
\usepackage{ulem}
\usepackage{fullpage}
\usepackage[obeyspaces,spaces]{url}
\usepackage{hyperref}
\usepackage{times}

\begin{document}

\date{April 2020}

\title{Heterogeneous CPU+GPU Stochastic Gradient Descent Algorithms}

\author{Yujing Ma and Florin Rusu\\
\{yma33, frusu\}@ucmerced.edu\\
University of California Merced
}

\maketitle

\begin{abstract}

The widely-adopted practice is to train deep learning models with specialized hardware accelerators, e.g., GPUs or TPUs, due to their superior performance on linear algebra operations. However, this strategy does not employ effectively the extensive CPU and memory resources -- which are used only for preprocessing, data transfer, and scheduling -- available by default on the accelerated servers. In this paper, we study training algorithms for deep learning on heterogeneous CPU+GPU architectures. Our two-fold objective -- maximize convergence rate and resource utilization simultaneously -- makes the problem challenging. In order to allow for a principled exploration of the design space, we first introduce a generic deep learning framework that exploits the difference in computational power and memory hierarchy between CPU and GPU through asynchronous message passing. Based on insights gained through experimentation with the framework, we design two heterogeneous asynchronous stochastic gradient descent (SGD) algorithms. The first algorithm -- CPU+GPU Hogbatch -- combines small batches on CPU with large batches on GPU in order to maximize the utilization of both resources. However, this generates an unbalanced model update distribution which hinders the statistical convergence. The second algorithm -- Adaptive Hogbatch -- assigns batches with continuously evolving size based on the relative speed of CPU and GPU. This balances the model updates ratio at the expense of a customizable decrease in utilization. We show that the implementation of these algorithms in the proposed CPU+GPU framework achieves both faster convergence and higher resource utilization than TensorFlow on several real datasets and on two computing architectures---an on-premises server and a cloud instance.

\end{abstract}

\section{INTRODUCTION}\label{sec:intro}

Deep learning has become a disruptive classification technology applied in a wide variety of domains, ranging from image~\cite{cvpr-sgd} and speech~\cite{speech-sgd} recognition to finance~\cite{finance-sgd} and combustion engines~\cite{combustion-dnn}. Building accurate deep learning models is expensive because the training process involves highly-intensive computations, e.g., the multiplication of large matrices. In order to speed up the process, the widely-adopted practice is to use specialized hardware accelerators, e.g., GPUs~\cite{facebook:large-batches} or TPUs~\cite{imagenet-minutes:iclr-2020}, due to their superior performance on linear algebra operations. CPU-only solutions require thousands of cores~\cite{imagenet-minutes:icpp-2018} to achieve similar performance---which is cost-ineffective. However, there is no real system composed only of accelerators---they are add-ons to standard architectures composed of CPUs and memory. In order to use the accelerator, data have to be preprocessed and passed through the system memory -- GPUDirect Storage~\cite{gpu-direct-storage} plans to avoid this with a direct data path between fast NVMe storage and GPU memory -- and the kernels have to be invoked. These procedures are coordinated by the CPU.

Based on the Amazon EC2 instances for accelerated computing~\cite{aws-ec2-gpu}, we observe a direct correlation between the number of GPUs -- on one side -- and the number of CPUs and the memory capacity---on the other. As a practical rule, there are 4-8 CPU cores and 30-60 GB of RAM for every GPU in the system. In addition to the Amazon EC2 instances, the CPU+GPU architecture is already part of some of the most powerful supercomputers on Top500~\cite{hpc-top500}, e.g., Summit~\cite{summit} and Titan~\cite{titan}. Due to their flexibility and high-performance, it is likely that CPU+GPU configurations will lead the way to exascale~\cite{heterogeneous:ieee-micro} through next-generation systems such as Perlmutter~\cite{perlmutter}. Thus, provisioning so many resources -- CPUs and memory -- only to preprocess data and schedule computation on the GPUs is wasteful from the deep learning perspective. Moreover, the number of CPU cores is continuously increasing. The latest CPUs released by Intel~\cite{intel-cpu} and AMD~\cite{amd-cpu} have 112 and 128 hardware threads, respectively. Similar to accelerators, this high degree of CPU parallelism can boost the linear algebra computations common in deep learning.

\paragraph*{Problem}
We study deep learning training on heterogeneous CPU+GPU architectures. Specifically, our objective is to design heterogeneous SGD algorithms that use efficiently all the resources available in a CPU+GPU system---not only a subpart. The main challenge consists in combining the characteristics of the two architectures -- the superior computational power of the GPU with the larger memory on the CPU -- into an SGD algorithm with optimal convergence behavior. While heterogeneous architectures have been used for model training before~\cite{tensorflow,omnivore}, this is done only in the context of the same SGD algorithm and considers how to optimally schedule data and computation across CPU and GPU. Our focus is on optimizing the interaction between the SGD algorithms performed on the two architectures---synchronous SGD on GPU and asynchronous SGD on CPU. The decision to consider architecture-specific algorithms is motivated by their theoretical~\cite{sync-vs-asynch-sgd,asynch-sgd-delay-comp} and empirical~\cite{bgd-vs-sgd:ipdps-2019} characteristics which recommend the use of synchronous SGD with large batches on GPU~\cite{facebook:large-batches} and asynchronous Hogwild SGD on CPU~\cite{hogwild,hogbatch,buckwild,hogwild-disk}.

\paragraph*{Contributions}
We introduce an adaptive framework for deep learning on heterogeneous CPU+GPU architectures that maximizes the utilization of each component during the entire execution. We achieve this by concurrent asynchronous coordination, dynamic data partitioning, and architecture-optimized algorithms. CPUs and GPUs are continuously assigned tasks -- which they perform concurrently -- by a lightweight asynchronous coordinator. The amount of data assigned to a task is dynamically and adaptively determined at runtime based on the current execution state. The CPU+GPU framework is generic and supports the implementation of most existing SGD algorithms~\cite{sgd:survey}. It is an invaluable testbed to evaluate existing algorithms and develop new ones.

We design two heterogeneous SGD algorithms with adaptive batch sizes. They are derived from the scalable asynchronous Hogbatch algorithm~\cite{hogbatch}. The first algorithm -- CPU+GPU Hogbatch -- combines small batches on CPU with large batches on GPU, which are both used to update a single shared model asynchronously. While providing better convergence than the single-architecture optimal algorithms, CPU+GPU Hogbatch hinders statistical convergence. The second algorithm -- Adaptive Hogbatch -- continuously monitors the number of updates performed by every task and changes the batch size dynamically based on the relative speed of CPU and GPU. This balances the model updates ratio at the expense of a customizable reduction in resource utilization.

We implement the two algorithms -- together with mini-batch and Hogbatch solutions for CPU and GPU-only -- in the heterogeneous CPU+GPU framework. We perform extensive experiments on several deep nets with increasing structural complexity over multiple real datasets. We execute the experiments both on an on-premises server at UC Merced, as well as on an AWS instance in the cloud. The results show that both algorithms achieve the fastest time to convergence and maximize the CPU and GPU utilization. Moreover, the heterogeneous algorithms outperform TensorFlow -- which performs similarly to our GPU-only algorithm -- by a significant margin.

\paragraph*{Outline}
The paper is organized as follows. We begin with a discussion of related work on architecture-optimized SGD algorithms in Section~\ref{sec:rel-work}. Preliminaries on SGD training for deep learning and heterogeneous CPU+GPU computing architectures are introduced in Section~\ref{sec:deep-learn} and Section~\ref{sec:architecture}, respectively. Our novel contributions are presented in Section~\ref{sec:framework} -- the framework for training on heterogeneous CPU+GPU architectures -- and Section~\ref{sec:adaptive-sgd}---the heterogeneous SGD algorithms. The experimental evaluation follows in Section~\ref{sec:experiments}, while Section~\ref{sec:conclusions} concludes the paper.

\section{RELATED WORK}\label{sec:rel-work}

We provide a comparison between the proposed framework and two other classes of systems that support deep learning on heterogeneous CPU+GPU architectures---TensorFlow and Omnivore. TensorFlow~\cite{tensorflow} -- and all the other related systems -- use heterogeneity at a smaller granularity. That is, they schedule linear algebra primitives across CPU and GPU. The decision on where to perform a primitive depends on the estimated execution time for each device. Unlike our framework, TensorFlow executes a single instance of the SGD algorithm which updates the unique model synchronously. There are a few problems with this approach. The amount of overlap between CPU and GPU execution is somewhat limited by the sequential structure of the DNN. Since the primitives have order dependencies, it is difficult to schedule more than one at a time. This results in the utilization of a single resource. Scheduling is heavily constrained by previous decisions because switching between CPU and GPU introduces time-consuming data transfers. Moreover, scheduling primitives instead of the complete SGD has more overhead. Similar to our framework, Omnivore~\cite{omnivore} splits the training data into batches having size proportional with the speed of the device. However, this size is statically computed and kept constant over the entire execution. The goal is to have perfectly synchronized execution with no delay across devices. The problem is that the actual speed at runtime can be quite different from the estimated one. We address this issue with dynamic batch sizes and asynchronous model updates. Heterogeneity is also considered in the distributed parameter server setting~\cite{hetero-ps}. The main difference from the centralized CPU+GPU architecture is that training data are statically partitioned to workers. Moving data between workers incurs expensive network traffic and is not viable. Instead, the applied solution uses different learning rates across workers. Similar to our work, the learning rate is computed based on the number of model updates. However, learning rate maintenance is more complex than modifying the batch size.

Although the relationship between the batch size and learning rate -- on one side -- and the number of updates, convergence, and utilization -- on the other -- is well-known~\cite{facebook:large-batches,crossbow}, there is an ongoing debate about the optimal batch size -- small or large -- and learning rate. Small batches generate more model updates, thus faster convergence. However, they do not saturate the high GPU throughput and result in low utilization. A practical solution is to increase the learning rate proportionally to the batch size~\cite{facebook:large-batches}. While this increases utilization, it also introduces convergence instability---especially close to the minimum. Our novel approach is to combine small and large batches in a single asynchronous SGD algorithm. The CPU performs a large number of small updates which move the loss function closer to the minimum faster. However, since they are based on a crude estimation of the gradient, they can be quite noisy. This is where the more accurate GPU updates are important---they move the loss in a better direction. Abstractly, we can think of the CPU updates as many small steps in a guessed direction, while the GPU updates are rare jumps using a compass. This combination of updates -- albeit sequential -- is theoretically proven to enhance SGD convergence~\cite{bertsekas:igd} and is at the origin of the SVRG family of algorithms~\cite{flex-ps}. We show empirically that it also improves convergence.

\section{SGD FOR DEEP LEARNING}\label{sec:deep-learn}

\begin{figure}[htbp]
\centering
\includegraphics[width=.7\textwidth]{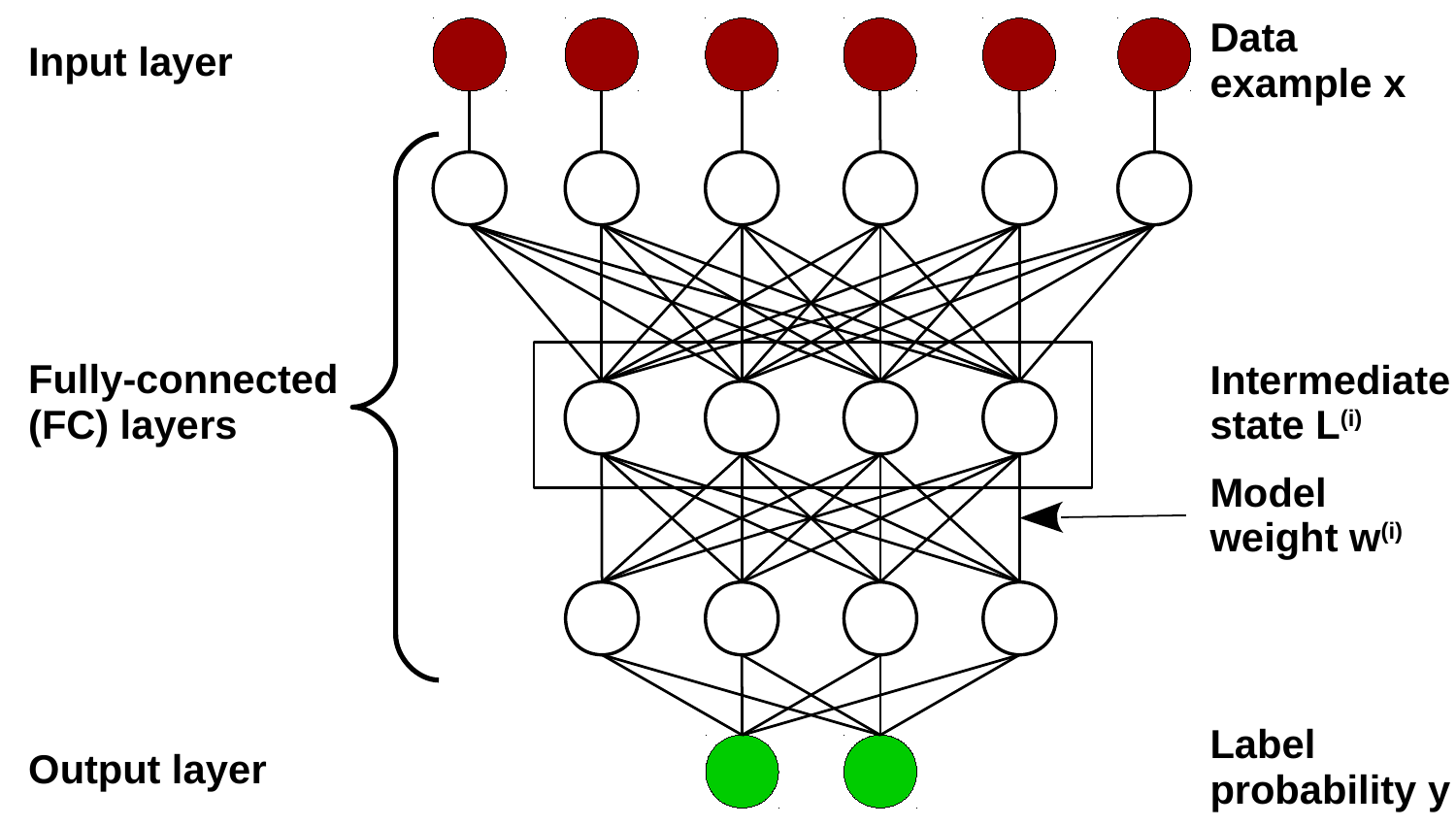}
\caption{DNN with three fully-connected (FC) layers.}
\label{fig:deep-net}
\end{figure}

The central component of deep learning is a \textit{Deep Neural Network (DNN)}~\cite{deep-nets-sgd}. As depicted in Figure~\ref{fig:deep-net}, a DNN is a layered network that takes as input an example given as a feature vector -- the input layer -- and produces the probability this example belongs to each class in a predefined set---the output layer. The intermediate layers are hidden to the user---they represent the model to be learned. Each layer contains a set of nodes or vertices. Nodes from two adjacent layers are connected by edges having weights and form a bipartite graph. If the graph is complete, i.e., there is an edge between any pair of nodes, the layer is called fully-connected (FC).

Formally, let the input data be a 2-D matrix $X \in \mathbb{R}^{N\times d_{1}}$ consisting of $d_{1}$-dimensional vectors $x_{i}$ for each of the $N$ examples. The vectors $x_{i}$ propagate through the DNN layers to the output, where their corresponding output labels $y_{i}$ are produced. Let the intermediate state of $x_{i}$ at layer $l$ be $L_{i}^{l}$, with $L_{i}^{1}=x_{i}$, where $L_{i}^{l}$ is a $d_{l}$-dimensional vector---the input changes its shape through the DNN. In each layer $l$, a series of linear algebra operations are applied to $L_{i}^{l}$ in order to generate $L_{i}^{l+1}$. The most intensive such operation is the matrix-vector product between vector $L_{i}^{l}$ and matrix $W^{l} \in \mathbb{R}^{d_{l+1}\times d_{l}}$ corresponding to the weights on the edges between the nodes in layers $l$ and $(l+1)$---the other operations are element-wise. If the DNN has $P$ layers and we denote the operation at layer $l$ by $\mathcal{F}_{l}$, then the complete processing of $x_{i}$ can be expressed as:
\begin{equation}\label{eq:forward}
y_{i} = \mathcal{F}_{P} \left(W^{P-1} \cdot \mathcal{F}_{P-1} \left(\dots \mathcal{F}_{1} \left( W^{1} \cdot x_{i} \right) \dots \right) \right)
\end{equation}
where $L_{i}^{l+1} = \mathcal{F}_{l} \left(W^{l} \cdot L_{i}^{l}\right)$ are the separate intermediate states. Essentially, a DNN is a composite function of embedded sub-functions over matrix-vector products between data and layer weights. DNN training corresponds to finding the optimal values for the weights in matrices $W^{l}$, $1\leq l\leq P$ -- denoted collectively the \textit{model} $W = \left\{ W^{1}, W^{2}, \dots, W^{P} \right\}$ -- that minimize the \textit{loss function} $\ell(X, W, Y)$ for the training dataset $X$---lower loss indicates high prediction accuracy. Here, $Y$ represents the set of known labels which are combined in the loss $\ell$ with the predictions corresponding to a fixed $W$.

\textit{SGD} is the most common method to train DNN models~\cite{bottou:optim-methods-scale-ml}. At high-level, SGD iteratively computes the gradient -- or derivative -- of the loss function over the training dataset and moves the model $W$ in the opposite direction of the gradient---which results in a decrease of the loss. Gradient computation requires a sequence of two passes over the DNN. In the \textit{forward pass}, the predicted labels are computed for the training data based on the current model $W$---the first model is randomly initialized. The \textit{backward pass} implements the chain rule of calculus for computing the gradient of a composite function starting from the predicted label $y_{i}$. If we denote by $\nabla\mathcal{F}_{l}$ the gradient with respect to model $W$ at layer $l$, then the back-propagation rule that computes the gradient $g_{i}$ is:
\begin{equation}\label{eq:backward}
g_{i} = \nabla\mathcal{F}_{1} \left( \dots \nabla\mathcal{F}_{P-1} \left( \nabla\mathcal{F}_{P} \left( \ell(y_{i}) \cdot W^{P} \right) \cdot W^{P-1} \right) \dots \right)
\end{equation}
We observe that the form of the forward and backward expressions in Eq.~(\ref{eq:forward}) and Eq.~(\ref{eq:backward}) are quite similar, having matrix-vector product as their dominant operation. The update equation at layer $l$ is:
\begin{equation}\label{eq:sgd-update}
W^{l} \longleftarrow W^{l} - \eta \cdot g_{i}^{l}
\end{equation}
where the \textit{learning rate $\eta$} is the scaling factor applied to the magnitude of the gradient. The learning rate is a hyperparameter of SGD---not a parameter of the DNN model. SGD can be stopped either after a fixed number of iterations, i.e., \textit{epochs}, or when there is no significant drop in the loss across iterations. In practice, due to the large dataset size and number of iterations it takes to converge, each SGD iteration is performed only over a randomly selected \textit{batch of $B$ training examples} -- not the entire dataset -- where $B$ is another hyperparameter. In this case, the matrix-vector multiplications become matrix-matrix multiplications, which are computationally more intensive, thus, the extensive use of GPUs in DNN training.

\begin{figure*}[htbp]
\centering
\includegraphics[width=\textwidth]{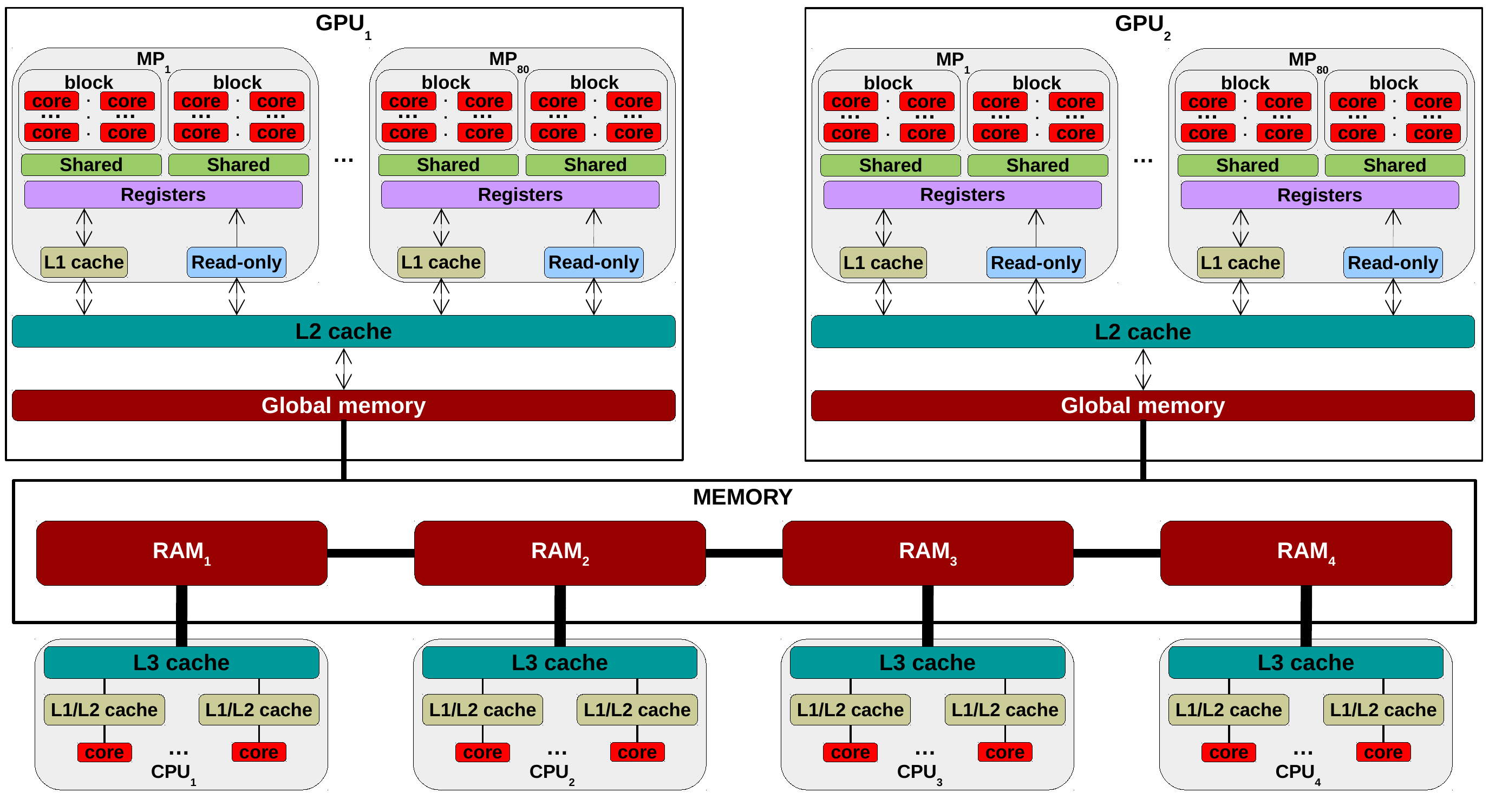}
\caption{Heterogeneous CPU+GPU architecture.}
\label{fig:hetero-arch}
\end{figure*}

\section{CPU+GPU ARCHITECTURE}\label{sec:architecture}

Figure~\ref{fig:hetero-arch} depicts graphically a heterogeneous CPU+GPU architecture with 4 CPUs and 2 GPUs connected together to the shared memory bus. Each CPU contains multiple cores and cache layers. The L1 and L2 caches are associated with each core, while the L3 cache is shared across all the cores in a CPU node. Each CPU is directly connected to a region of the DRAM memory. The CPUs are connected to each other by high-bandwidth interconnects. To access DRAM regions on other nodes, data is transferred over these interconnects. However, this is slower than accessing the local memory, thus, the non-uniform memory access (NUMA) pattern. NUMA cache-coherency is implemented in hardware, thus implicit.

A GPU contains multiple streaming multiprocessors (MP). Each MP consists of a large number of specialized cores targeted at a limited subset of instructions. In the CUDA programming model, work is issued to the GPU in the form of a function, referred to as the kernel. A logical instance of the kernel executed on an MP core is called a thread. The kernel code is parametrized by a logical thread identifier that allows each thread to operate on a different partition of the input data---which has to be moved explicitly between the CPU and GPU memory. Since thousands of threads can be executed concurrently across MPs, global thread synchronization is not available. Nonetheless, synchronization can be enforced at thread block level. Threads can access the various units of the deep memory hierarchy in Figure~\ref{fig:hetero-arch} explicitly in the code. When a global memory address is requested by a thread, aligned successive addresses are converted into a single memory transaction---memory coalescing. Thus, consecutive threads have to access consecutive addresses in order to minimize the number of memory transactions.

\begin{table}[htbp]
  \centering
	\resizebox{\textwidth}{!}
	{
	\begin{tabular}{l||rr|rr}
		& \multicolumn{2}{c|}{\textbf{UC Merced}} & \multicolumn{2}{c}{\textbf{AWS p3.16xlarge}} \\
		& \multicolumn{1}{c}{CPU} & Tesla K80 GPU & \multicolumn{1}{c}{CPU} & Volta V100 GPU \\
    	\hline
		cores & 14 & 192 per MP & 18 & 172 per MP \\
		blocks & --- & 16 per MP & --- & 32 per MP \\
		threads & 28 & 2048 per MP & 36 & 2048 per MP \\
		L1 cache & 32(I) + 32(D) KB & 48 KB & 32(I) + 32(D) KB & 128 KB \\
		L2 cache & 256 KB & 1.5 MB & 256 KB & 6 MB \\
		L3 cache / shared memory & 35 MB & 48 KB & 45 MB & 96 KB \\
		MEMORY / global memory & 256 GB & 12 GB & 488 GB & 16 GB \\
	    \hline
   	\end{tabular}
    }  
    \caption{Hardware architecture specifications.}\label{tbl:hardware-spec}
\end{table}

Table~\ref{tbl:hardware-spec} gives the hardware specification of two servers with CPU+GPU architecture used throughout this work. The UC Merced server is an on-premises machine with two CPUs and an NVIDIA Tesla K80 GPU. The other server is an AWS p3.16xlarge cloud instance with two CPUs and the latest Volta V100 GPU. While the number of cores and threads is much larger for the GPUs, the numbers for the CPUs are also quite high, e.g., 28 and 36 independent threads, respectively, can run concurrently on a single CPU. Since both systems have two CPUs, there are 56 and 72 CPU threads overall. However, this is not sufficient to reduce the gap in performance given by the superior GPU degree of parallelism---there are 80 MPs on the V100 GPU. Although the amount of memory available on the CPU is 20X to 30X larger than on the GPU, the L2 cache on the GPU is 6X to 24X larger. This reflects the throughput emphasis of the GPU memory hierarchy as opposed to the latency optimization for CPU.

\section{DNN FRAMEWORK ON CPU+GPU}\label{sec:framework}

We pursue two main objectives in designing the CPU+GPU framework for deep learning. First, the framework is a generic testbed to evaluate existing SGD algorithms and develop new ones. This is achieved by a modular architecture in which components are assigned independently to hardware resources. An SGD algorithm is expressed by a series of primitive operations and a communication strategy between components. The second objective is to maximize the utilization of every resource during execution. We achieve this by concurrent asynchronous coordination, dynamic data partitioning, and architecture-optimized SGD algorithms. CPUs and GPUs perform concurrent asynchronous SGD algorithms -- specialized for their specific architecture -- on data assigned dynamically and adaptively at runtime based on the current execution state. In this section, we present the architecture and workflow of the proposed DNN framework on heterogeneous CPU+GPU architectures.

\subsection{Framework Architecture}\label{ssec:framework:architecture}

\begin{figure}[htbp]
\centering
\includegraphics[width=.75\textwidth]{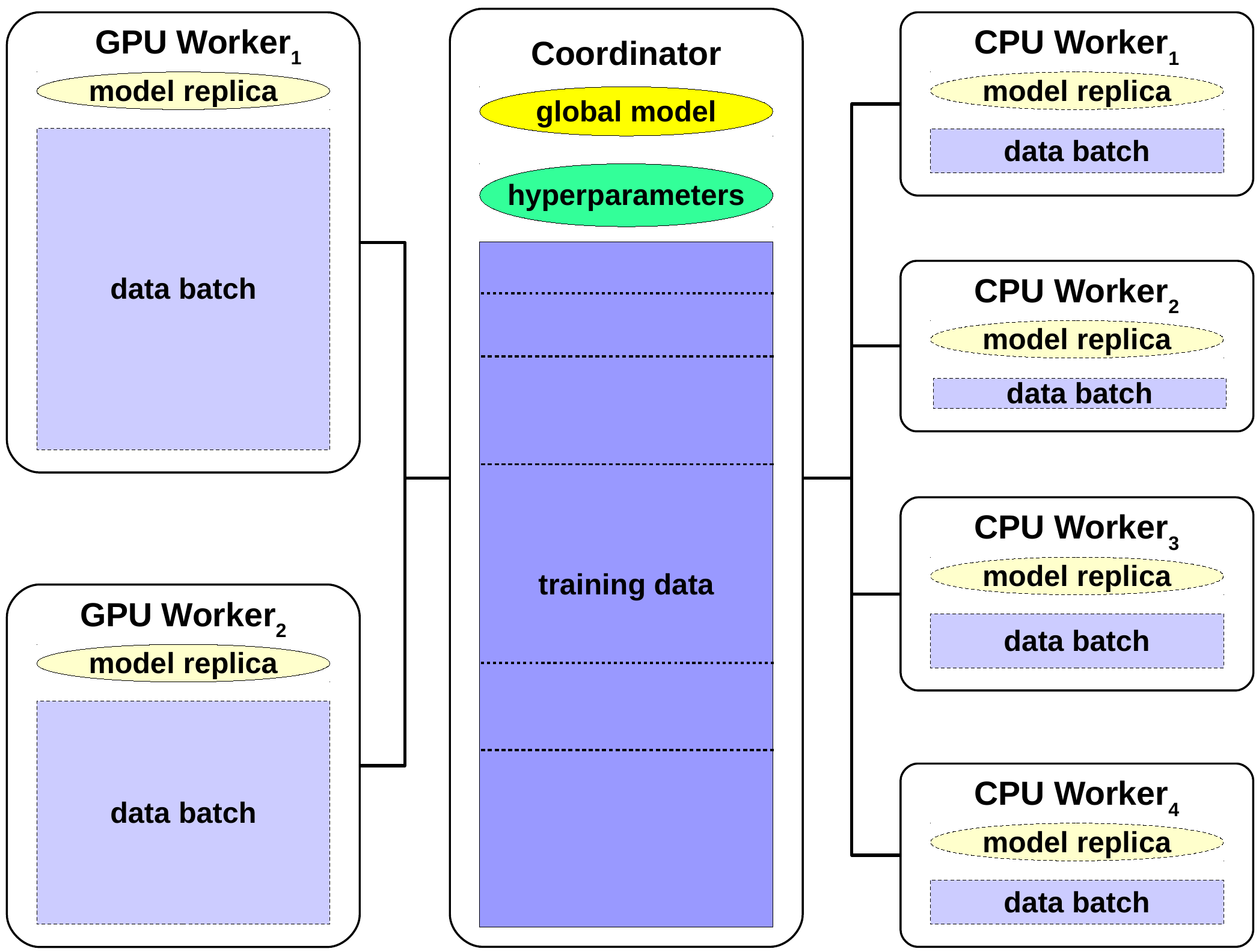}
\caption{Framework architecture.}
\label{fig:framework-arch}
\end{figure}

The architecture of the heterogeneous CPU+GPU framework for deep learning is depicted in Figure~\ref{fig:framework-arch}. It consists of a series of asynchronous worker threads corresponding to each of the CPUs and GPUs in Figure~\ref{fig:hetero-arch}, and a central coordinator. In this example, there are four CPU workers and two GPU workers. However, in a shared virtualized environment such as Amazon EC2, the framework can be assigned only a subset of the available hardware resources. The coordinator and workers are implemented as stand-alone system threads that exist over the entire duration of the program. The worker assigned to a hardware component is in charge of managing the resources, e.g., cores, memory, threads, and operation of that component. The coordinator assigns data and tasks to workers, and schedules their interaction. The communication between the coordinator and workers -- workers do not communicate directly -- is realized through control messages, while data are passed through references in the shared memory space. In the case of deep learning, these data include the model -- and its gradient -- and the training examples split into batches. The coordinator maintains the global model and prepares the training data. Each worker is assigned a model replica -- which can be either a deep or shallow copy of the global model -- and a data batch---which is a reference to a range in the training data at the coordinator. Handling the training data is simpler because it requires read-only access. The hyperparameters of the SGD algorithm are maintained by the coordinator.

\paragraph*{Coordinator}
The coordinator corresponds to the parameter server in distributed~\cite{parameter-server} or multi-GPU~\cite{geeps,crossbow} settings. Its main role is to control workers' access to the global model through the model update policy. Since the coordinator thread processes messages sequentially, the default policy is synchronous model updates---the replicas are applied to the global model one after another, in the order in which they are received. In order to support asynchronous updates, the model update logic has to be moved to the workers. After computing the gradient, the workers apply it to their replica reference -- a pointer to the global model -- concurrently. In this case, the burden on the coordinator is considerably smaller because it does not execute any part of the SGD algorithm.

In our shared memory framework, the coordinator plays an additional role that is completely missing from distributed parameter servers~\cite{parameter-server,geeps}. The coordinator assigns data batches of different size dynamically and adaptively to the workers based on their processing speed. This is a fundamental feature in a heterogeneous CPU+GPU architecture. If the same batch size is given to a CPU and a GPU worker, the GPU worker would process a number of batches equal to the ratio between the speed of the GPU and that of the CPU~\cite{omnivore}. Since this ratio is significant, the GPU would process hundreds or thousands of batches while the CPU processes a single batch. The end result would be that the CPU updates are ignorable. Alternatively, if CPU and GPU are synchronized, the GPU would be stalled most of the time. In order to cope with this issue, the coordinator continuously monitors the number of updates each worker executes and changes the batch size such that the difference between the fastest and slowest worker is bounded. This strategy is implemented inside the model update procedure and requires only a simple reference assignment. As far as we know, this is the first learning framework that supports dynamic batch sizes across concurrent workers. In all the other solutions, the training data is statically partitioned and distributed to workers.

\paragraph*{CPU Workers}
The workers are statically associated with a computational resource -- CPU socket or GPU -- and perform an iteration of the SGD algorithm on the assigned batch and model. Since CPU workers share the same address space with the coordinator, they have direct access to the global model and training data. This allows for reference access and avoids deep copies---the dotted lines in Figure~\ref{fig:framework-arch}. However, due to the uneven NUMA memory access, references can introduce unexpected cache coherency effects~\cite{dimm-witted,hogbatch,hogwild-disk}. The CPU worker has to consider another level of parallelism -- corresponding to the local cores/threads -- when performing the SGD algorithm. The alternatives are to compute a single gradient over the entire batch or to split the batch into smaller sub-batches and compute a gradient for each. In the first case, intra-thread parallelism is applied only to the linear algebra operations and is encapsulated in the corresponding library functions, e.g., Intel MKL~\cite{intel-mkl}. In the second case, there are two levels of parallelism---inter-thread parallelism across sub-batches and intra-thread parallelism inside a sub-batch. The inter-thread parallelism has to be implemented in the CPU worker. This can be done with explicit threads or with higher-level constructs such as OpenMP~\cite{openmp}. Based on the level of parallelism and the model update policy, many variations of the SGD algorithm can be designed~\cite{dimm-witted}---supported by the framework with different implementations of the CPU worker.

\paragraph*{GPU Workers}
A GPU worker is associated with every GPU accelerator in the system---for which it serves as the exclusive interface. The GPU worker coordinates the memory transfers between CPU and GPU, and invokes kernel execution on the GPU---all these happen asynchronously and with minimal interference on the other system components. This allows for advanced GPU features, such as data transfer through the unified memory address space and kernel execution through asynchronous streams, to be isolated in the GPU worker. The execution of the SGD algorithm on GPU follows the standard pattern of first moving the data and the model, and then invoking kernels for the linear algebra operations, e.g., from the cublas library~\cite{cublas}, on the forward and backward DNN passes. By default, the intermediate output of kernel invocations is kept in the GPU memory in order to reduce data movement. However, advanced memory management strategies that work at layer granularity~\cite{memory-mgmt:ipdps-2019} can also be added. The main difference between the CPU and GPU worker is how they handle the model---the model replica in the GPU worker is always a deep copy of the global model. This is because the replica serves as a transition buffer between CPU and GPU, which is accessed only during transfers between the two. Multiple read/write accesses to the global model while moved in/out of the GPU memory may have unexpected consequences.

\begin{figure}[htbp]
\centering
\includegraphics[width=.8\textwidth]{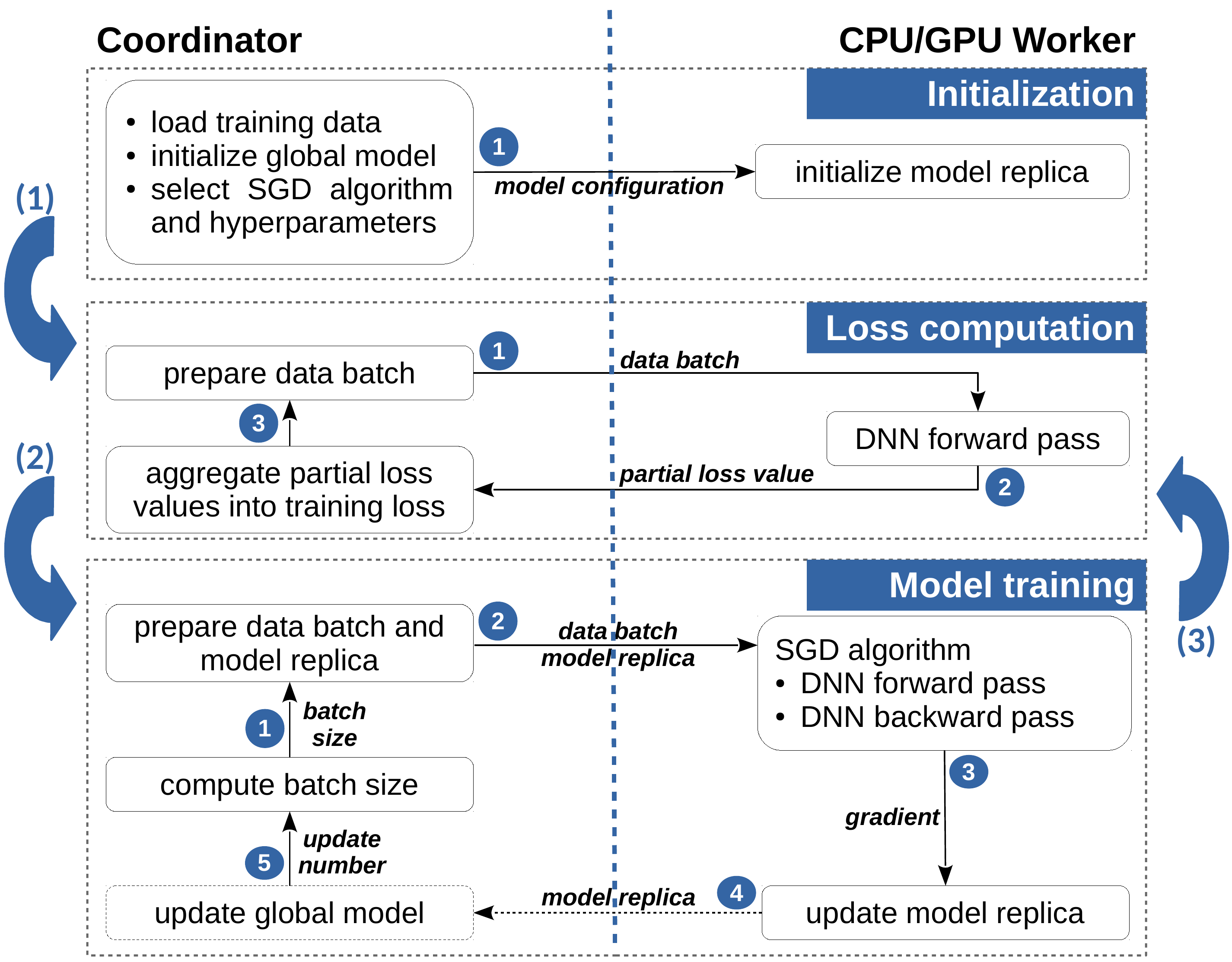}
\caption{Framework execution workflow.}
\label{fig:framework-workflow}
\end{figure}

\subsection{Framework Workflow}\label{ssec:framework:workflow}

Figure~\ref{fig:framework-workflow} illustrates how the framework performs the SGD algorithm for deep learning. The tasks executed by each worker, as well as the messages exchanged with the coordinator, are shown in the figure. During initialization, the coordinator loads the training data in memory and prepares it for the linear algebra operations in SGD. The global model is allocated and initialized with arbitrary/random values. The model configuration is passed to the workers for their initialization of the model replicas. This is necessary only for the GPU workers which have to allocate memory on the device. Since multiple SGD implementations are supported, the coordinator has to select an algorithm and its hyperparameters for each worker, and a global model update policy. While these are currently specified by the programmer, we envision a solution in which the active workers and their algorithms are selected automatically based on the characteristics of the data and the model.

Although loss computation is not a required stage of the SGD algorithm, it is necessary to depict how the accuracy of the model evolves. As such, the framework performs loss computation after each complete pass -- or a given number of batches -- over the training data. The loss is computed with a DNN forward pass over the training -- or test -- data. The coordinator splits the data into batches which are assigned to workers. The size of the batch is proportional to the worker speed. If this is larger than the memory capacity of the worker, the batch is split further into sub-batches. Each worker computes a partial loss on its data batch and then sends it back to the coordinator, which aggregates it into the overall loss. This strategy is optimized for execution time by prioritizing the fast workers and minimizing the coordinator overhead.

The SGD algorithm is performed in the model training stage. At each iteration, the coordinator starts by determining the batch size corresponding to every worker. Initially, the size is proportional to the worker speed. Later, the size changes adaptively based on the number of model updates performed by the worker. The coordinator prepares a batch by selecting a continuous range from the training data and storing a reference to its starting position. The model replica is initialized with the current state of the global model. This can be a reference to the global model or a full deep copy. The batch and the replica are passed to the worker to execute the SGD algorithm selected in the initialization stage. This is the main part of model training and consists of a forward and backward pass through the DNN to compute the gradient. Finally, the gradient is applied to update the model---another DNN forward pass. If the update is applied to the deep replica, this has to be subsequently integrated in the global model---which can be done synchronously at the coordinator or asynchronously at the worker. In the case of reference replicas, the update is directly applied to the global model. The last step in model training -- which closes the loop -- is the message sent by the worker to the coordinator to inform that the update has been applied. Since these messages are processed sequentially, the next batch size is computed individually for each worker based on its number of updates.

\section{SGD WITH ADAPTIVE BATCH SIZES}\label{sec:adaptive-sgd}

The CPU+GPU framework supports the implementation of heterogeneous versions of most -- if not all -- existing SGD algorithms~\cite{sgd:survey}. Modifications are confined only to the type of messages exchanged between coordinator and the workers, and how they are handled. In this section, we introduce two heterogeneous SGD algorithms derived from the scalable asynchronous Hogbatch~\cite{hogbatch}. We choose Hogbatch as our base SGD algorithm because of two reasons. First, it supports asynchronous updates. These are perfectly suited for the speed difference between CPU and GPU. Second, unlike Hogwild~\cite{hogwild}, Hogbatch operates on batches. This matches better the highly-parallel GPU architecture optimized for throughput---the larger the batch, the higher the utilization~\cite{crossbow}. The experimental results in Section~\ref{sec:experiments} confirm the necessity to design these specialized algorithms---and their superiority over standard Hogbatch.

\begin{algorithm}[htbp]
\caption{Hogbatch}\label{alg:hogbatch}

\begin{minipage}[t]{.48\textwidth}
\textit{\textbf{Coordinator: ScheduleWork Message Handler}}
\algsetup{linenodelimiter=.}

\begin{algorithmic}[1]

\REQUIRE ~~\\
Worker $\mathcal{E}$ asking for work\\
Set of batches $\mathbb{B}$ with $b$ training examples

\IF {$\mathbb{B} \neq \emptyset$}
\STATE Extract next batch $\mathcal{B}$ from $\mathbb{B}$
\STATE $\mathbb{B} \leftarrow \mathbb{B} \smallsetminus \mathcal{B}$
\STATE Send message \textit{ExecuteWork ($\mathcal{B}$)} to worker $\mathcal{E}$
\ENDIF

\end{algorithmic}
\end{minipage}
\hfill
\begin{minipage}[t]{.5\textwidth}

\textit{\textbf{Worker $\mathcal{E}$: ExecuteWork Message Handler}}
\algsetup{linenodelimiter=.}

\begin{algorithmic}[1]

\REQUIRE ~~\\
Batch $\mathcal{B}$ with $b$ training examples

\STATE Gradient: $g \leftarrow \nabla\mathcal{F}_{1} \left( \dots \nabla\mathcal{F}_{P} \left( \ell(\mathcal{B}) \cdot W^{P} \right) \dots \right)$

\STATE Update model: $W \leftarrow W - \eta \cdot g$

\STATE Send message \textit{ScheduleWork($\mathcal{E}$)} to coordinator

\end{algorithmic}
\end{minipage}
\hfill
\end{algorithm}

\subsection{Hogbatch}\label{ssec:sgd:hogbatch}

The mapping of Hogbatch to our framework is given in Algorithm~\ref{alg:hogbatch}. The main task of the coordinator is to serve work requests from workers. For this, the coordinator extracts a batch of $b$ training examples and sends them to the requesting worker. It is important to notice that \textit{the same batch size $b$ is given to all the workers}. When there are no more batches and all the workers are done, an SGD epoch has finished and the process is restarted with the full training dataset. While the coordinator executes requests serially, the workers process batches concurrently. First, they compute the gradient of the assigned batch on the current DNN model. Then, they update the model with the computed gradient. In Hogbatch, the DNN model is shared across all the workers---the local replicas are references to the global model. Since the workers read and modify the model concurrently without any synchronization primitives, conflicts are unavoidable. However, the speedup provided by parallel processing outweighs the impact of update conflicts and results in faster overall convergence~\cite{hogwild,hogbatch}.

\subsection{CPU+GPU Hogbatch}\label{ssec:sgd:fixed-hogbatch}

The direct application of Hogbatch in a heterogeneous CPU+GPU architecture raises two problems. First, the model has to be copied to the GPU memory, thus, access by reference does not work. Our solution is to create a deep copy of the global model in the GPU worker \textit{ExecuteWork} message handler. The GPU kernels operate exclusively on this replica. Once the replica is updated, we push it to the global model asynchronously. If the GPU workers have similar speed, they perform a similar number of updates and their local replicas do not become stale. Otherwise, merging a local stale replica requires careful consideration. A possible solution is to perform model update on the current global model, which is copied again to the GPU~\cite{hogwild-disk}. In this case, the gradient is computed on a model, while the update is performed on another -- most recent -- model. Additionally, the learning rate can be decreased to compensate for the stale gradient~\cite{hetero-ps}, diminishing the importance of the update.

The second problem is triggered by using the same batch size $b$ across all the workers. Due to the orders of magnitude difference in processing speed between CPU and GPU, the GPU performs considerably more updates. In the worst case, the CPU takes more time to process a single batch than the GPU processing all the others. This behavior is detrimental because the CPU ends up doing useless work and, moreover, it stalls the GPU. Our novel solution is to \textit{use different batch sizes for CPU and GPU}. The CPU batch size is set to $t$ -- where $t$ is the number of cores or threads on the CPU -- so that each thread processes exactly one example. The rationale for this choice is to ensure that all the threads are utilized. This special case of Hogbatch is the Hogwild algorithm~\cite{hogwild}. The GPU batch size is set to a value that satisfies two conditions. First, it guarantees a high enough utilization of the GPU. Second, the time to process a batch on GPU is close to the time on CPU. However, the GPU memory capacity imposes an upper bound on the size. Based on these constraints, the GPU batch size varies from a few hundreds to several thousands, depending on the DNN structure. This idea can be generalized to having a different batch size for every worker. Notice that supporting different batch sizes across workers requires minimal changes to the \textit{ScheduleWork} message handler in Algorithm~\ref{alg:hogbatch}.

While the benefit of using different batch sizes is important to reduce staleness among workers, it may be argued that its impact on convergence is harder to assess. Indeed, there is no theoretical analysis for Hogbatch with different batch sizes. However, the analysis of any SGD asynchronous algorithm makes strong simplifying assumptions~\cite{hogwild} that rarely hold in practice. Intuitively, the interaction between small and large batches improves convergence because it combines a large number of model updates based on inaccurate gradients -- corresponding to small batches -- with updates from accurate gradients computed on large batches. This idea represents the principle of the entire family of SVRG algorithms~\cite{flex-ps} which are theoretically proven to have asymptotically better convergence. Our conjecture -- supported empirically -- is that convergence remains superior even when the two types of updates are applied concurrently. Moreover, we set the learning rate to be proportional with the batch size~\cite{facebook:large-batches}---we have both different batch sizes and different learning rates. This guarantees that the impact of the more accurate gradients on convergence is higher.

\begin{algorithm}[htbp]
\caption{Adaptive Hogbatch}\label{alg:adaptive-hogbatch}

\begin{minipage}[t]{.53\textwidth}
\textit{\textbf{Coordinator: ScheduleWork Message Handler}}
\algsetup{linenodelimiter=.}

\begin{algorithmic}[1]

\REQUIRE ~~\\
Worker $\mathcal{E}$ asking for work;
Number of model updates $u^{\mathcal{E}}$ performed by worker $\mathcal{E}$;
Batch size $b^{\mathcal{E}}$ for worker $\mathcal{E}$;
Minimum ($\textit{min}_{u}$) and maximum ($\textit{max}_{u}$) number of updates performed by all other workers except $\mathcal{E}$;
Minimum ($\textit{min}^{\mathcal{E}}_{b}$) and maximum ($\textit{max}^{\mathcal{E}}_{b}$) batch size threshold for worker $\mathcal{E}$;
Training dataset $\mathbb{B}$

\item[\textit{$\triangleright$ Update batch size $b^{\mathcal{E}}$ for worker $\mathcal{E}$}]

\IF {$u^{\mathcal{E}} < \textit{min}_{u}$}
\STATE $b^{\mathcal{E}} \leftarrow \textit{maximum} \left( b^{\mathcal{E}}/\alpha, \textit{min}^{\mathcal{E}}_{b} \right)$; $\textit{min}_{u} \leftarrow u^{\mathcal{E}}$
\ELSIF {$u^{\mathcal{E}} > \textit{max}_{u}$}
\STATE $b^{\mathcal{E}} \leftarrow \textit{minimum} \left( b^{\mathcal{E}} \cdot \alpha, \textit{max}^{\mathcal{E}}_{b} \right)$; $\textit{max}_{u} \leftarrow u^{\mathcal{E}}$
\ENDIF

\item[\textit{$\triangleright$ Prepare and send batch to worker $\mathcal{E}$}]

\IF {$b^{\mathcal{E}} \leq |\mathbb{B}|$}
\STATE Extract batch $\mathcal{B}$ of size $b^{\mathcal{E}}$ from $\mathbb{B}$
\STATE $\mathbb{B} \leftarrow \mathbb{B} \smallsetminus \mathcal{B}$
\STATE Send message \textit{ExecuteWork ($\mathcal{B}$)} to worker $\mathcal{E}$
\ENDIF

\end{algorithmic}
\end{minipage}
\hfill
\begin{minipage}[t]{.46\textwidth}

\textit{\textbf{CPU Worker $\mathcal{E}$: ExecuteWork Message Handler}}
\algsetup{linenodelimiter=.}

\begin{algorithmic}[1]

\REQUIRE ~~\\
Batch $\mathcal{B}$ with $b$ training examples\\
Number of threads $t$

\STATE Split $\mathcal{B}$ into $t$ sub-batches $\{\mathcal{B}_{1},\dots,\mathcal{B}_{t}\}$ of size $\mathcal{B}/t$

\STATE \textbf{for} $i=1$ \textbf{to} $t$ \textbf{do \textit{in parallel}}
\STATE \hspace*{.25cm} Gradient:\\
\hspace*{.25cm} $g_{i} \leftarrow \nabla\mathcal{F}_{1} \left( \dots \nabla\mathcal{F}_{P} \left( \ell(\mathcal{B}_{i}) \cdot W^{P} \right) \dots \right)$
\STATE \hspace*{.25cm} Update model: $W \leftarrow W - \eta \cdot g_{i}$
\STATE \textbf{end for}

\STATE $u^{\mathcal{E}} \leftarrow u^{\mathcal{E}} + t \cdot \beta$ \textit{$\triangleright$ Increase number of model updates}
\STATE Send message \textit{ScheduleWork($\mathcal{E}$,$u^{\mathcal{E}}$)} to coordinator

\end{algorithmic}
\end{minipage}
\end{algorithm}

\subsection{Adaptive Hogbatch}\label{ssec:sgd:adaptive-hogbatch}

The problem with CPU+GPU Hogbatch is that the batch sizes have to be determined prior to execution. This can be a lengthy trial-and-error process that adds complexity to hyperparameter tuning. Moreover, the batch sizes are static and they do not take into consideration the runtime execution environment. This can lead to unbounded divergence between the number of updates across CPU and GPU, which manifests by loss function instability and, ultimately, slower convergence.

We address these issues in Adaptive Hogbatch---depicted in Algorithm~\ref{alg:adaptive-hogbatch}. The main idea is to continuously monitor the workers' status and \textit{update the batch size dynamically based on the number of updates}. This can be done in the \textit{ScheduleWork} message handler at the coordinator. While the relationship between the number of updates and resource utilization is clear, the connection to convergence is not straightforward, especially when the updates are computed over batches with different size. The number of updates has to be large. Due to computational and memory constraints, this can be achieved only with small batches. However, small batches generate inaccurate gradients---which hurt convergence. In order to address these conflicting goals, we apply two criteria when computing the batch size. First, the gap in the number of updates between the slowest and fastest worker has to be bounded. This is achieved by slowing down (i.e., increasing the batch size) the worker with the largest number of updates or speeding up (i.e., decreasing the batch size) the worker with the smallest number of updates, respectively. The value of the batch size is scaled up or down by a constant factor $\alpha$ which is a user-defined parameter set by default to $2$---the batch size is doubled or halved, respectively. The second goal is to maintain a minimum level of utilization on every worker. For this, we define lower and upper thresholds on the batch size, which we do not allow to be crossed. Alternatively, we can monitor the actual utilization for devices that provide such APIs. The initial batch size is set to the lower threshold for CPU and the upper threshold for GPU. The computation of a new batch size is light and does not incur observable overhead at the coordinator.

The CPU worker in Adaptive Hogbatch (Algorithm~\ref{alg:adaptive-hogbatch}) has to maintain the number of model updates it performs. This poses some complications because of the asynchrony incurred by the nested Hogbatch execution. While $t$ threads perform updates, these are conflicting, thus, it is not clear how many survive. We quantify this uncertainty through the user-defined parameter $\beta$ which specifies the fraction of surviving updates. When $\beta=1$ -- the default value determined empirically -- the CPU worker performs $t$ updates per batch. The closer $\beta$ gets to $0$, the fewer updates are considered by the coordinator when computing the new batch size.

\section{EXPERIMENTAL EVALUATION}\label{sec:experiments}

The purpose of the experimental evaluation is to investigate the following questions:
\begin{itemize}[leftmargin=*,noitemsep,nolistsep]
\item Are the heterogeneous Hogbatch algorithms improving upon the CPU and GPU-only alternatives in time to convergence and statistical efficiency?
\item How does the heterogeneous framework compare with the state-of-the-art TensorFlow?
\item What is the impact of the computing architecture on the algorithms in general? Particularly, how is the GPU impacting the performance?
\item What is the ratio of model updates among CPU and GPU?
\item What utilization do the Hogbatch algorithm implementations in our framework achieve on CPU and GPU?
\end{itemize}

\subsection{Setup}\label{sec:experiments:setup}

\paragraph*{Implementation}
We implement the heterogeneous CPU+GPU framework for deep learning in C/C++ using the pthreads library. The coordinator and each worker is managed by a stand-alone thread. The threads communicate using our custom asynchronous message queue. The CPU worker schedules Hogbatch instances on its corresponding cores using dynamic OpenMP (3.7.0-3) threads. The linear algebra operations on CPU are implemented with Intel MKL (2.187) functions invoked inside OpenMP threads. The GPU worker invokes kernels written in CUDA 10.0 which call the linear algebra primitives from the Nvidia cublas (10.2.1.243-1) library. The TensorFlow (1.13.1) implementation consists of the driver program in which the DNN architecture and the objective function are defined~\cite{slide-code}. Then, the mini-batch SGD optimizer is called to perform the training. All the code is available online as open source~\cite{cpu+gpu-code}.

\paragraph*{Hardware}
We perform the experiments on two systems---a fully-managed on-premises UC Merced server and the Amazon AWS p3.16xlarge instance~\cite{aws-ec2-gpu}. The specification of these two computing architectures is given in Table~\ref{tbl:hardware-spec}. The UC Merced server has 56 CPU threads and an Nvidia Tesla K80 GPU running on Ubuntu 16.04 SMP with Linux kernel 4.4.0-98. We assign 48 out of the 56 threads to a single CPU worker in order to simplify their use in OpenMP. This allows up to 48 threads to perform concurrent model updates. The number of OpenMP threads for linear algebra operations is also set to 48. The remaining threads are the stand-alone workers and coordinator. Since the Tesla K80 GPU consists of two independent GPUs programmed independently, we allocate a separate GPU worker to each of them. The Amazon AWS p3.16xlarge instance has 8 Nvidia Volta V100 GPUs and 64 vCPUs or threads---AWS limits the number of threads available on an instance. We use the standard AWS configuration from March 2020. We assign 56 out of the 64 threads to a single CPU worker for up to 56 concurrent model updates, while the number of OpenMP threads is set to 60. In order to maximize GPU utilization, we run experiments with a single GPU in this case. In TensorFlow, the default settings for the GPU version are employed. These default to using a single GPU, independent of the number of available GPUs. Thus, while our framework uses two GPUs on the UC Merced server, TensorFlow uses only one.

\begin{table*}[hbpt]
	\centering
	\resizebox{\textwidth}{!}{
    \begin{tabular}{l|rrrr|rr|r}
    \textbf{dataset} & \textbf{\#examples} & \textbf{\#features} & \textbf{\#labels} & \textbf{size} & \textbf{CPU batch size} & \textbf{GPU batch size} & \textbf{DNN architecture} \\
    \hline
		covtype & 581,012 & 54 & 2 &485 MB & [1-64] & [128-8,192] & 54-512-512-512-512-512-512-2 \\
		w8a & 64,700 & 300 & 2 & 155 MB & [1-64] & [64-8,192] & 300-512-512-512-512-512-512-512-512-2 \\
    delicious & 16,105 & 500 & 983 & 128 MB & [1-32] & [64-2,048] & 500-512-512-512-512-512-512-512-512-983 \\ 
		real-sim & 72,309 & 20,958 & 2 & 12.1 GB & [1-64] & [64-8,192] & 20,958-512-512-512-512-2 \\
		\hline
    \end{tabular}
	}    
 
	\caption{Experimental datasets. CPU and GPU batch size are the minimum and maximum batch size used in the Hogbatch algorithms. DNN architecture corresponds to the layers and the number of units per layer.}\label{tbl:datasets}
\end{table*}

\paragraph*{Datasets and DNN configurations}
We consider four real data sets -- depicted in Table~\ref{tbl:datasets} -- that exhibit large variety in size, features, and number of classes. These datasets have been used previously to evaluate the performance of parallel SGD on CPU and GPU~\cite{Xclassification-dataset-repo,bgd-vs-sgd:ipdps-2019}---more details can be found therein. We process all the datasets in dense format. The batch size on CPU varies between 1-64 examples per thread, while for GPU it ranges between 64-8,192. The number of hidden layers is set inversely proportional to the dataset size---to 4 (real-sim), 6 (covtype), and 8 (w8a and delicious). The number of units in a hidden layer is kept constant at 512. Since all the layers are fully-connected, the processing complexity is proportional to the number of layers.

\paragraph*{Methodology}
We execute each algorithm for the same fixed amount of time. This is chosen such that the loss converges for at least one algorithm. The minimum loss across all the algorithms is taken as basis for comparison. All the loss values are normalized to this basis. This process measures which algorithm converges fastest to a certain loss---the ultimate goal in practice. The upper and lower thresholds for batch size are varied with the datasets and the DNN architecture. The GPU utilization for the lower threshold is about 50\%, while for the upper threshold is close to 100\%. The initial batch size is set to the upper threshold on the GPU workers. The CPU worker starts with a batch size of 1 example per thread---it performs Hogwild. The number of model updates is measured as the average over all the epochs. All the algorithms are initialized with the same model, which gives the same initial loss. The initial values of the DNN weights are randomly drawn from a normal distribution with standard deviation equal to the number of units in the current layer. The sigmoid function is used as activation in the hidden layers. Softmax activation is applied to the output layer in order to compute the cross-entropy loss. The SGD learning rate is chosen by griding its range in powers of 10 and selecting the value that achieves the lowest loss across all the algorithms. The batch size and learning rate are correlated and set according to~\cite{facebook:large-batches}. We emphasize that the same hyperparameters are used for the same hardware architecture. The time to load the data, output the result, and evaluate the loss are not included in time measurements.

\begin{figure}[htbp]
\begin{center}
\subfloat[covtype K80]{\includegraphics[width=0.47\textwidth]{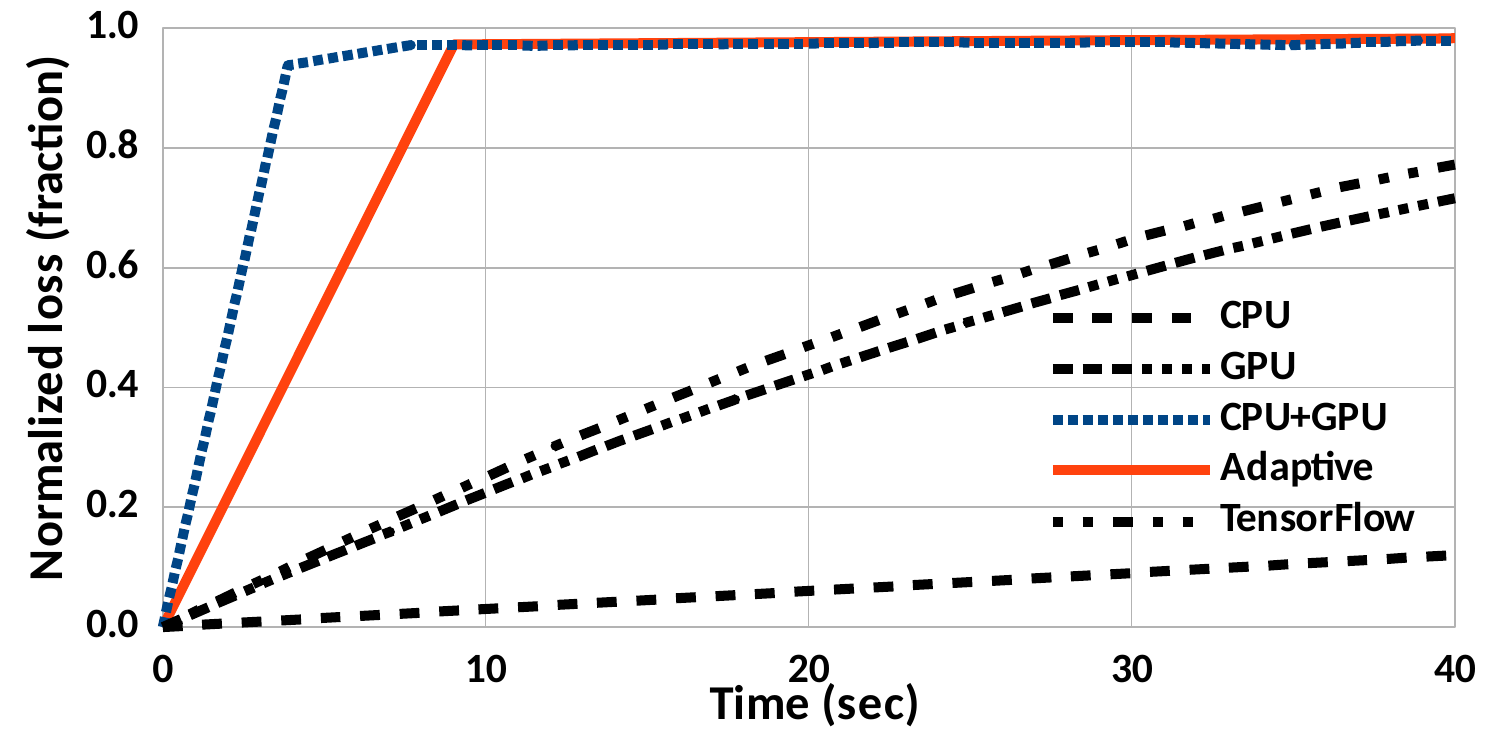}\label{fig:f-time}}\hfill
\subfloat[covtype V100]{\includegraphics[width=0.47\textwidth]{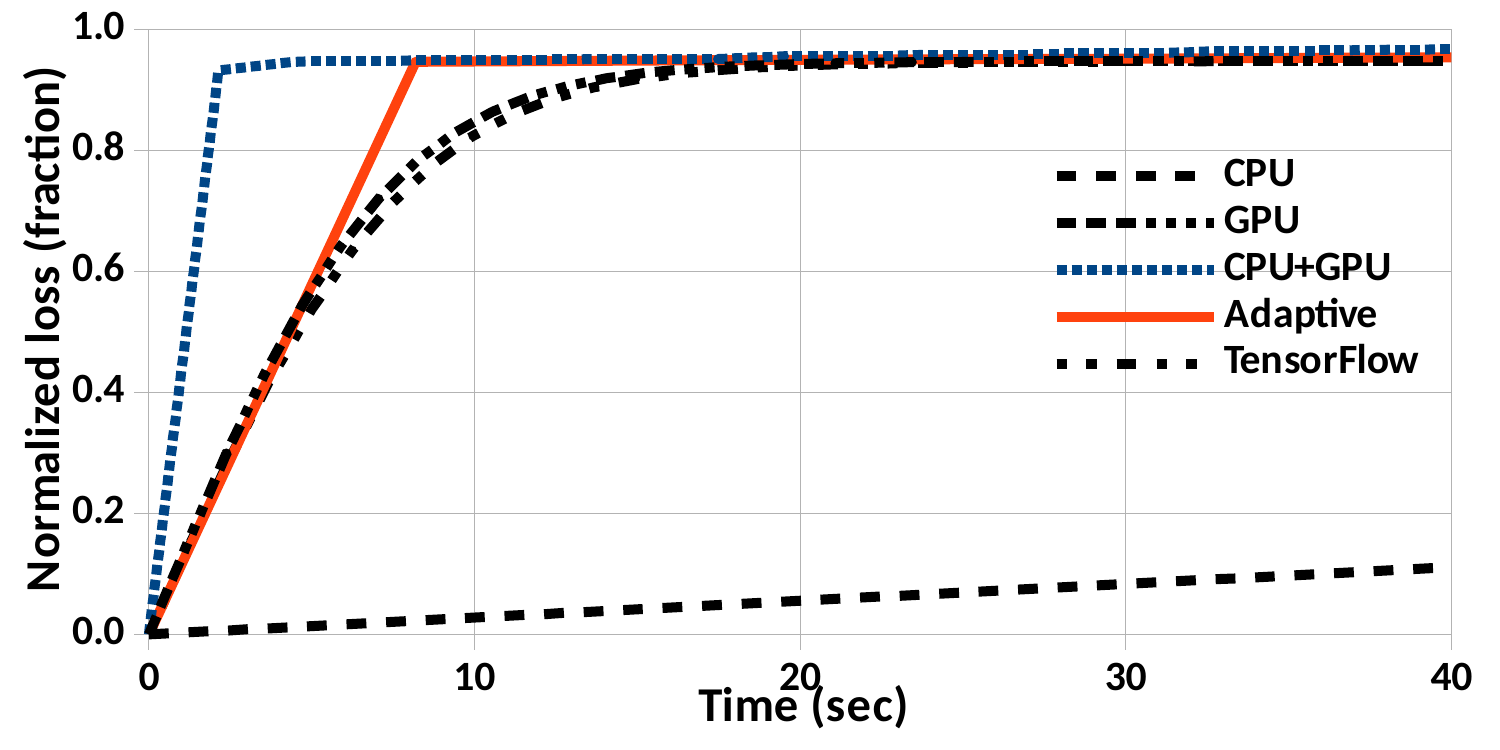}\label{fig:f-time-aws}} \\
\subfloat[w8a K80]{\includegraphics[width=0.47\textwidth]{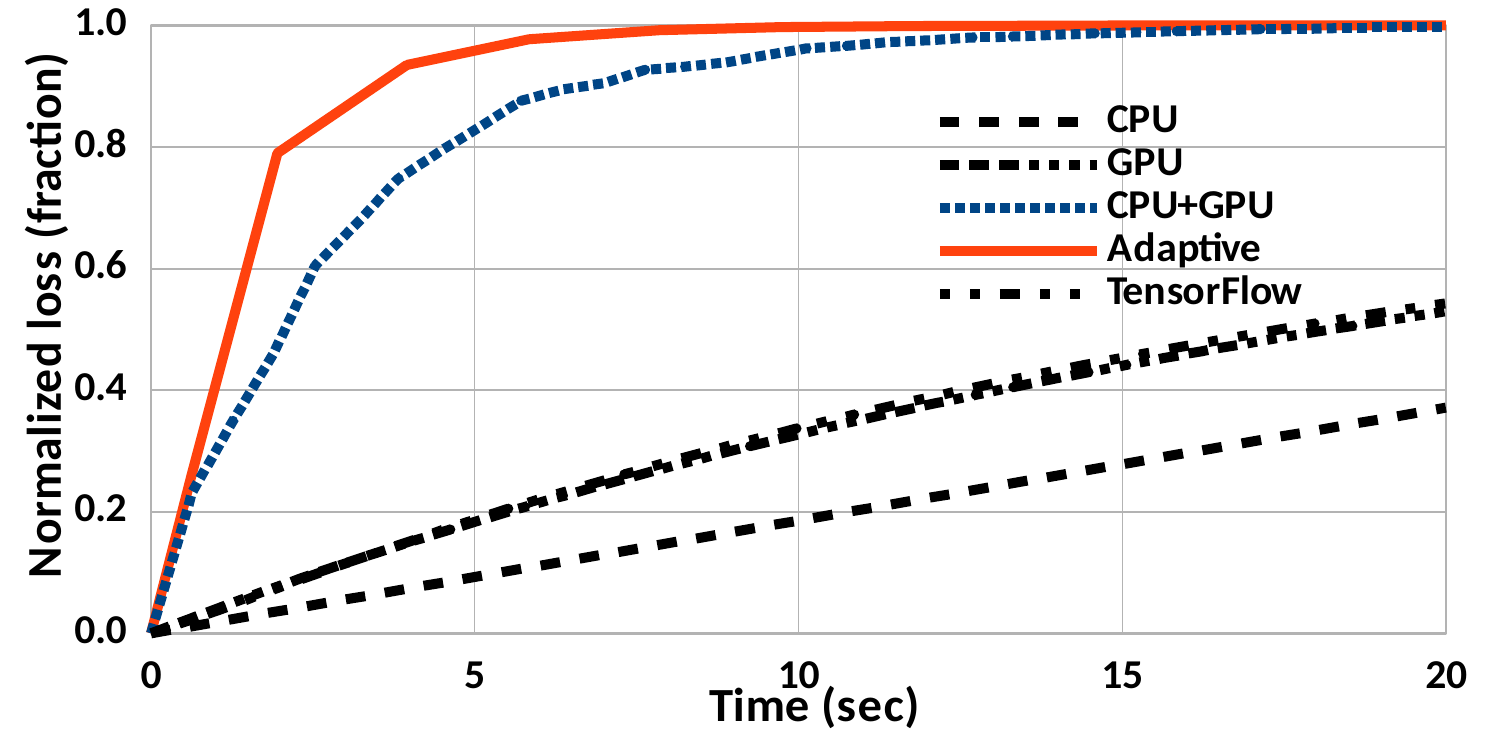}\label{fig:w-time}}\hfill
\subfloat[w8a V100]{\includegraphics[width=0.47\textwidth]{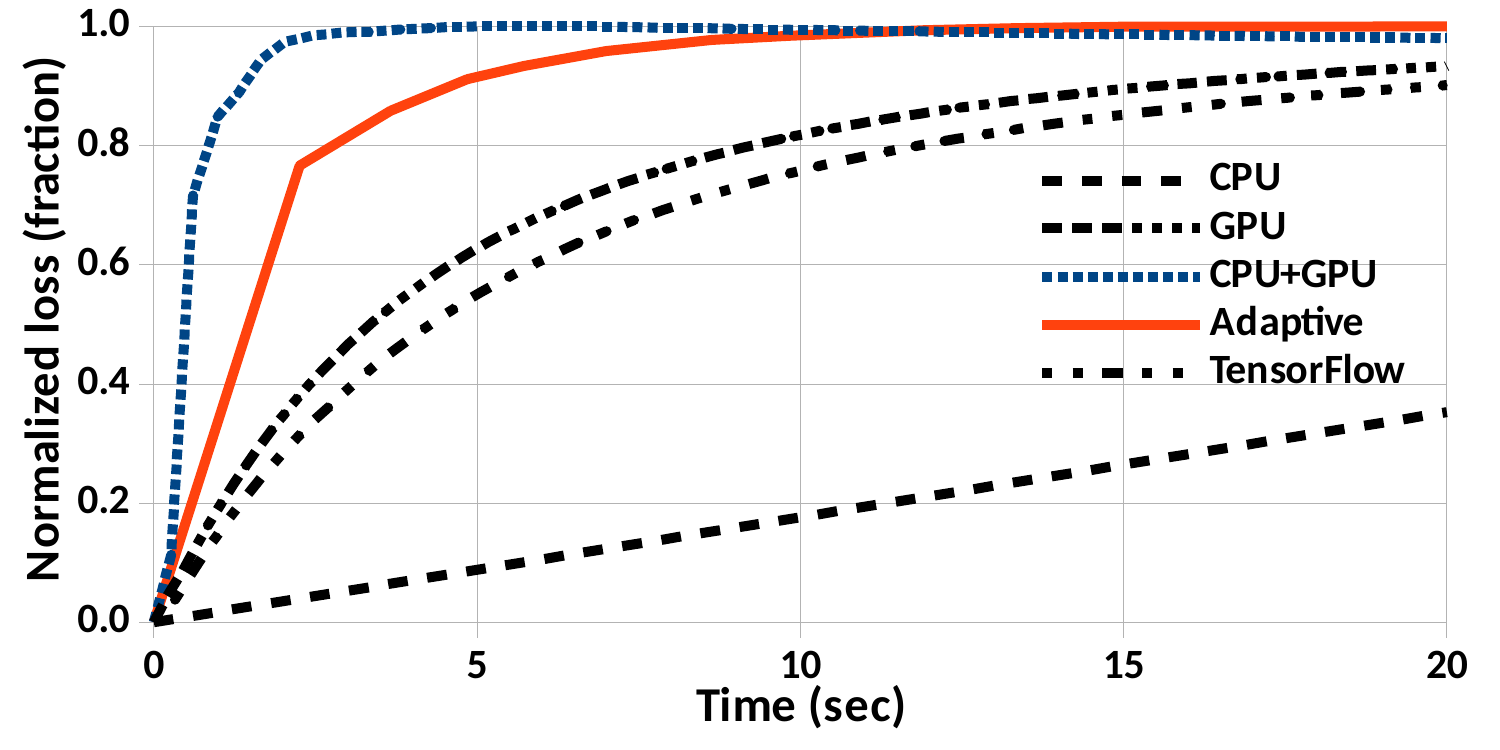}\label{fig:w-time-aws}} \\
\subfloat[delicious K80]{\includegraphics[width=0.47\textwidth]{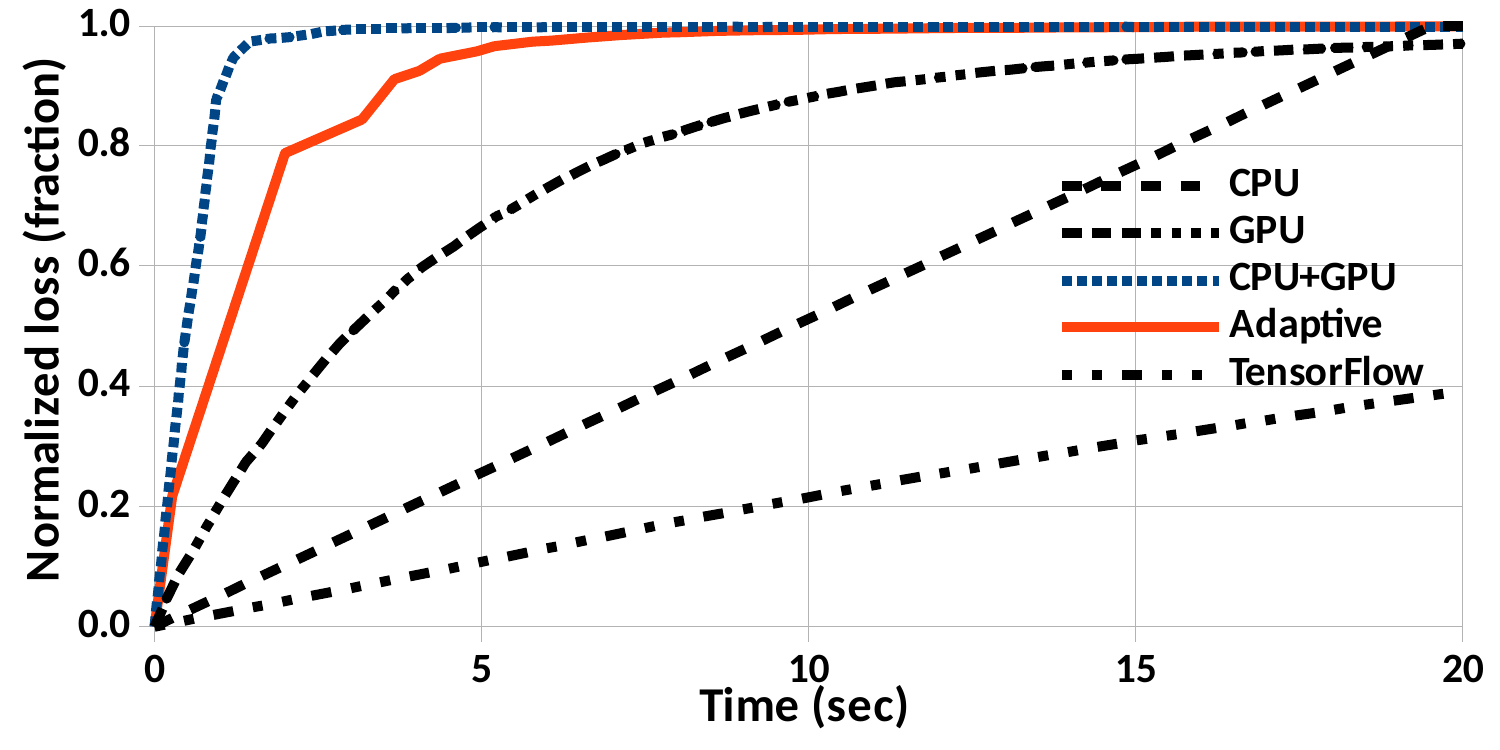}\label{fig:d-time}}\hfill
\subfloat[delicious V100]{\includegraphics[width=0.47\textwidth]{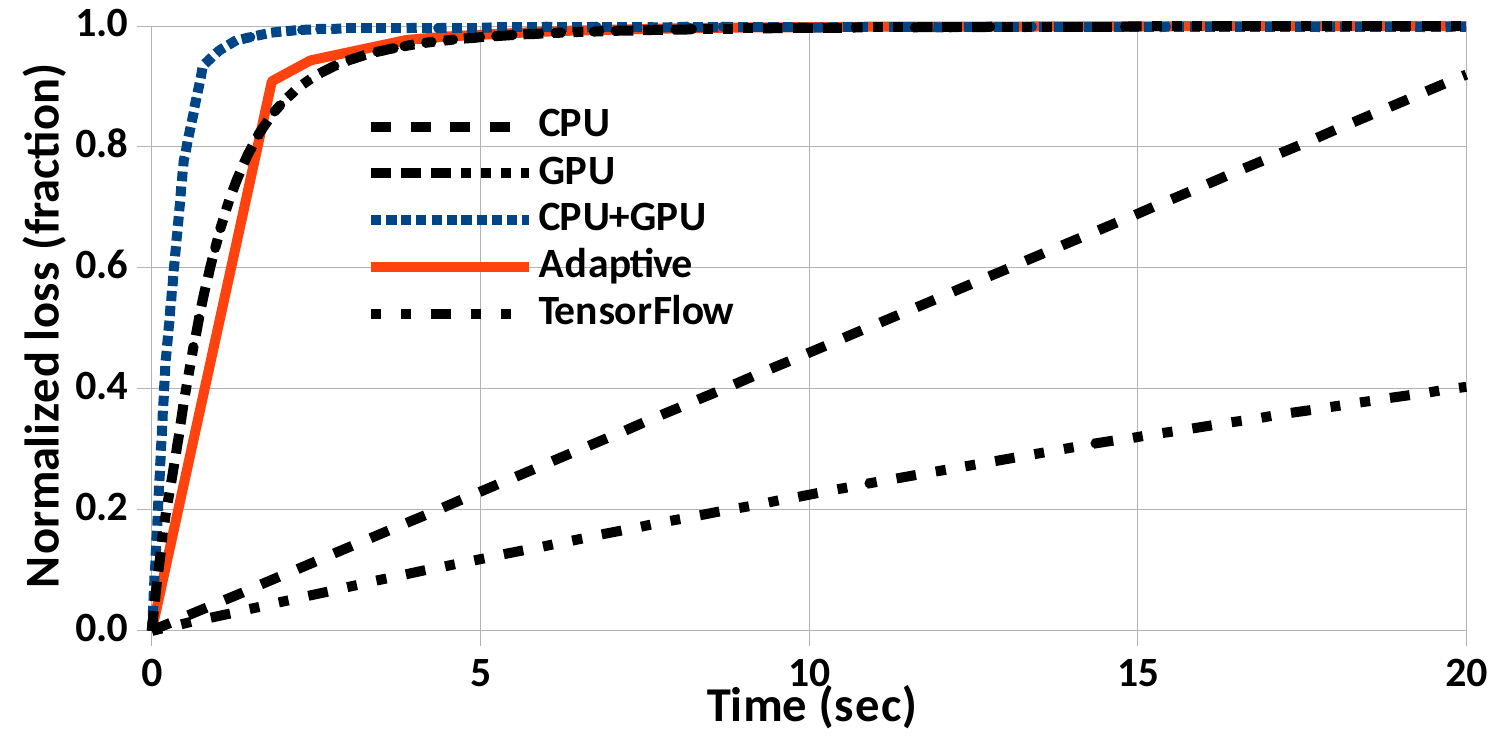}\label{fig:d-time-aws}} \\
\subfloat[real-sim K80]{\includegraphics[width=0.47\textwidth]{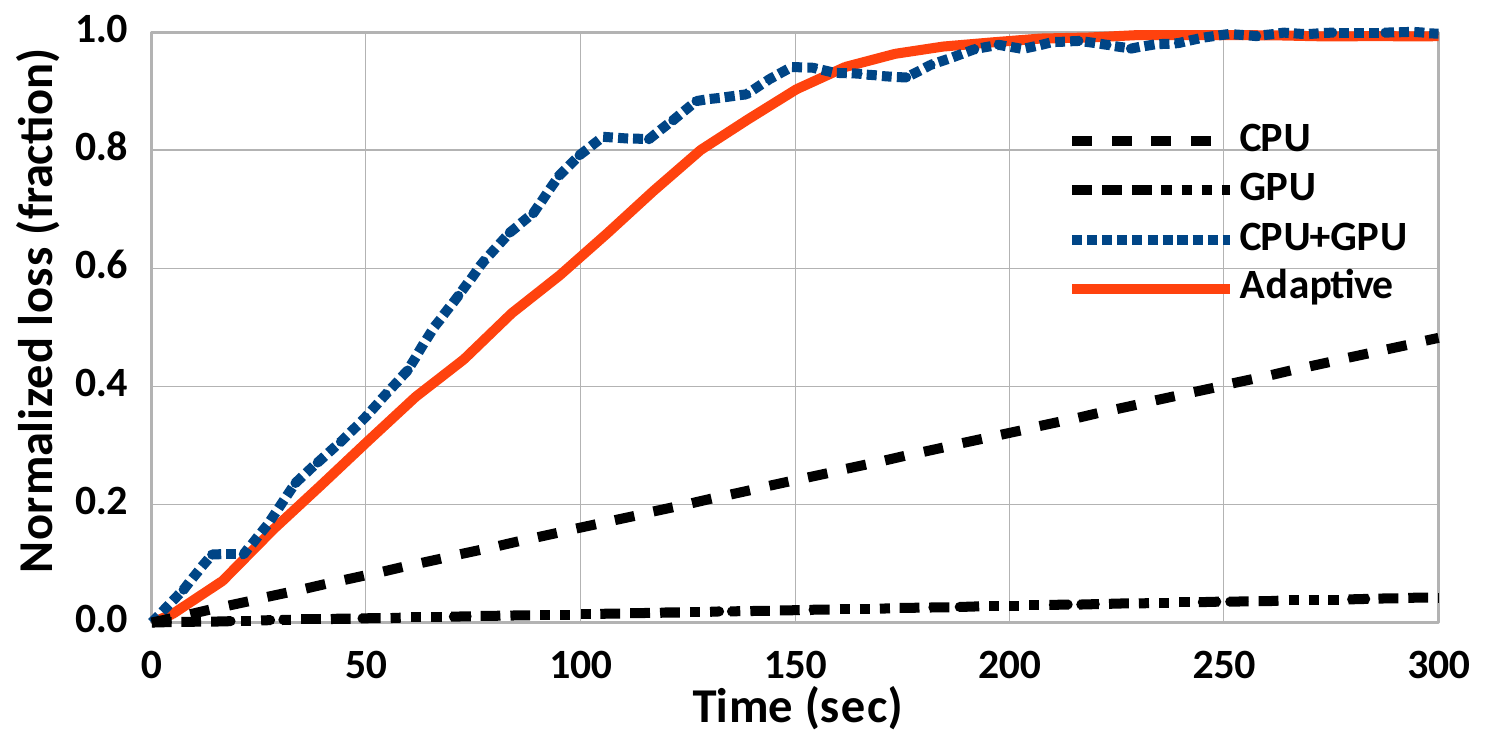}\label{fig:r-time}}\hfill
\subfloat[real-sim V100]{\includegraphics[width=0.47\textwidth]{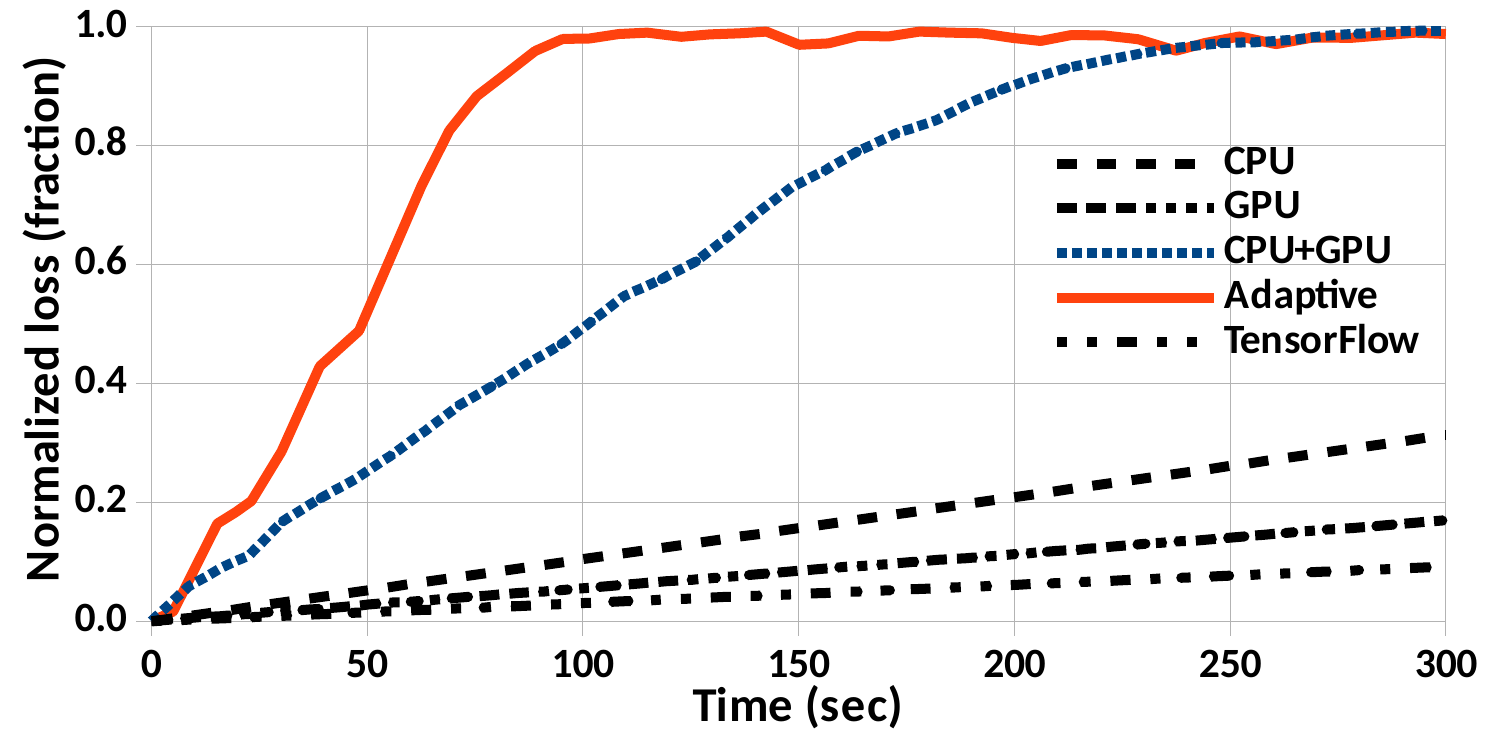}\label{fig:r-time-aws}}
\caption{Normalized loss for time to convergence on the UC Merced server (K80) and the AWS p3.16xlarge instance (V100). Left side is for UC Merced and right side is for AWS.}\label{fig:loss-as-time}
\end{center}
\end{figure}

\subsection{Results}\label{sec:experiments:results}
We include four Hogbatch algorithms -- CPU, GPU, CPU+GPU, and Adaptive -- and TensorFlow in the experiments. Hogbatch CPU is Hogwild performed on CPU-only, while Hogbatch GPU -- and TensorFlow -- are essentially mini-batch SGD on the AWS instance. On the UC Merced server, Hogbatch GPU performs two concurrent -- and asynchronous -- updates.

\paragraph*{Time to convergence}
The behavior of the normalized loss as a function of the elapsed time is depicted in Figure~\ref{fig:loss-as-time}---left side on the UC Merced server, right side on the AWS instance. In both cases, Hogwild CPU takes considerably longer -- from 236X to 317X -- to execute an SGD epoch than GPU, thus its loss follows a slope increasing at a much slower -- linear -- rate in the beginning. In fact, Hogwild CPU did not finish an epoch in the allocated time budget for any of the datasets---it got close only for delicious. Nonetheless, the number of model updates per epoch is the highest among all the methods. The relative performance between CPU and GPU matches perfectly the relationship between Hogwild (CPU) and (mini-) Hogbatch (GPU). On low-dimensional data, (mini-) Hogbatch converges faster. However, as the dimensionality increases, there is a switch between the two, with Hogwild clearly outperforming (mini-) Hogbatch on real-sim. As expected, TensorFlow mirrors almost identically the convergence curve of (mini-) Hogbatch (GPU). The only exception is delicious, on which TensorFlow has much worse convergence. The reason is the multi-label classification -- 983 vs. 2 labels -- which is much slower in TensorFlow. Notice also that TensorFlow crashes for the real-sim dataset on the UC Merced server due to insufficient memory. It is evident that the heterogeneous Hogbatch algorithms achieve the steepest decrease in loss per unit of time. The mixture of small and large batches combines the best behavior of the CPU and GPU solutions and improves upon them significantly in all the cases. While CPU+GPU outperforms Adaptive Hogbatch in more cases, Adaptive reaches the minimum loss in less than half of the time for w8a on UC Merced and real-sim on the AWS instance, respectively. This is because batch sizes having more uniform values, as dictated by the relative performance of CPU and GPU, generate fewer conflicts in these two cases. We conclude by pointing out that, while 90\% of the minimum loss is achieved fast, further improvement is rather slow on covtype.

\begin{figure}[htbp]
\begin{center}
\subfloat[covtype K80]{\includegraphics[width=0.47\textwidth]{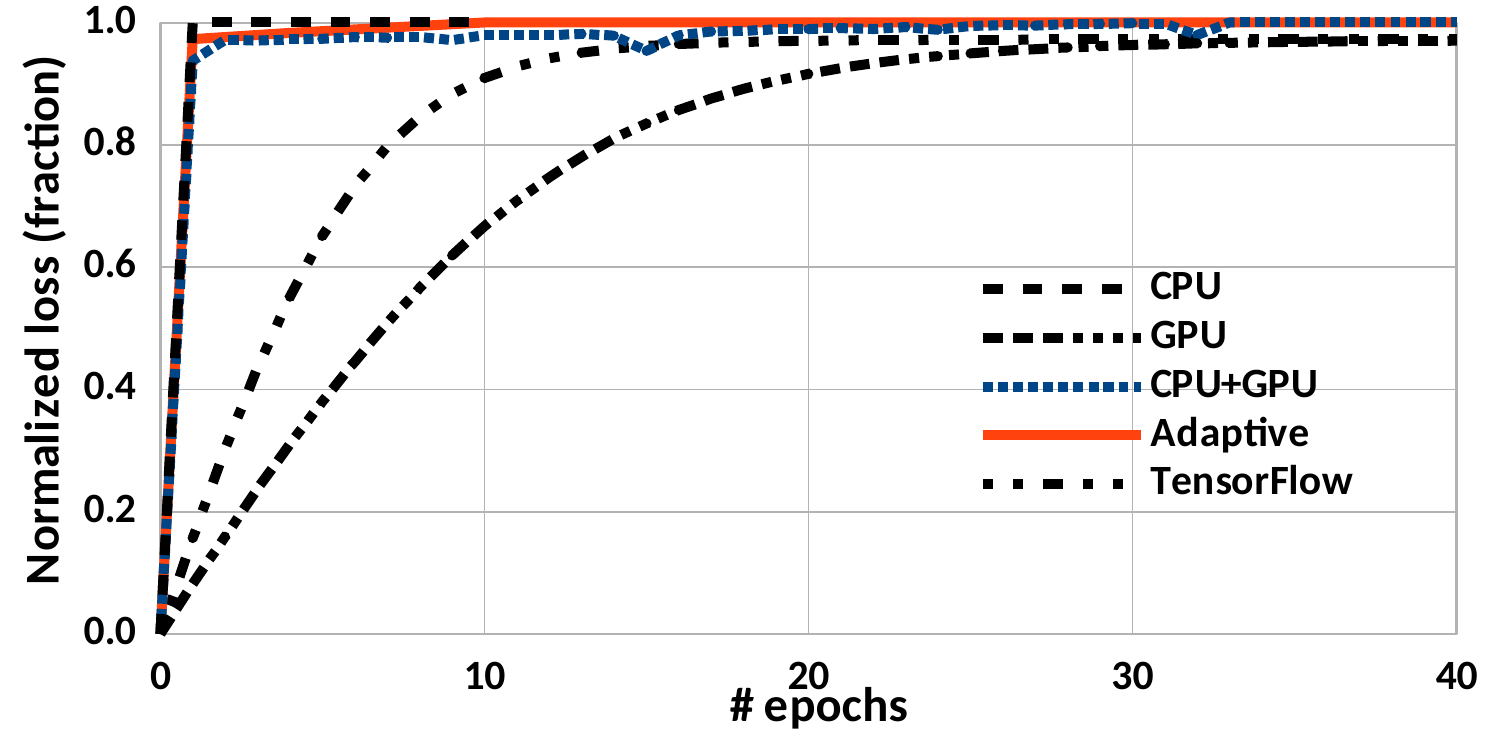}\label{fig:f-iter}}\hfill
\subfloat[covtype V100]{\includegraphics[width=0.47\textwidth]{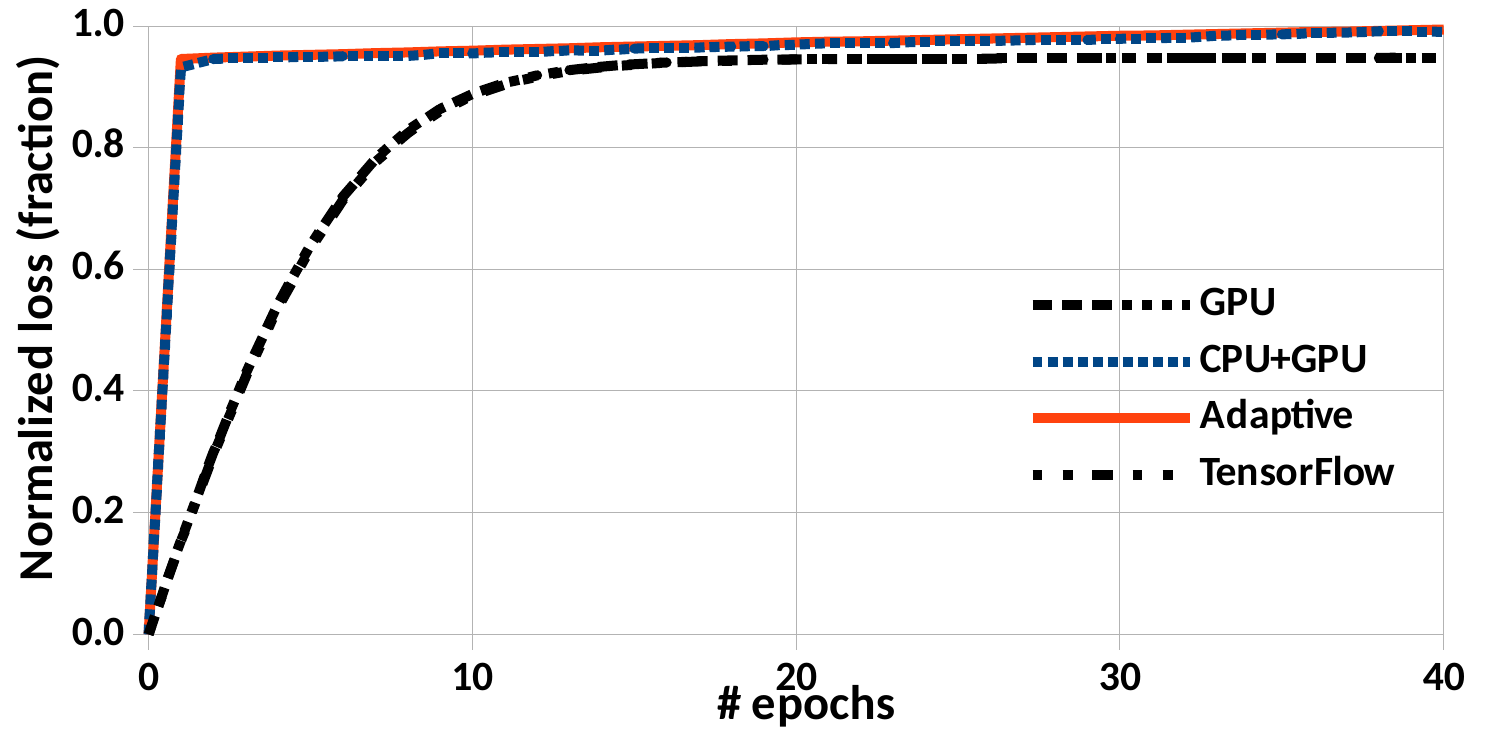}\label{fig:f-iter-aws}} \\
\subfloat[w8a K80]{\includegraphics[width=0.47\textwidth]{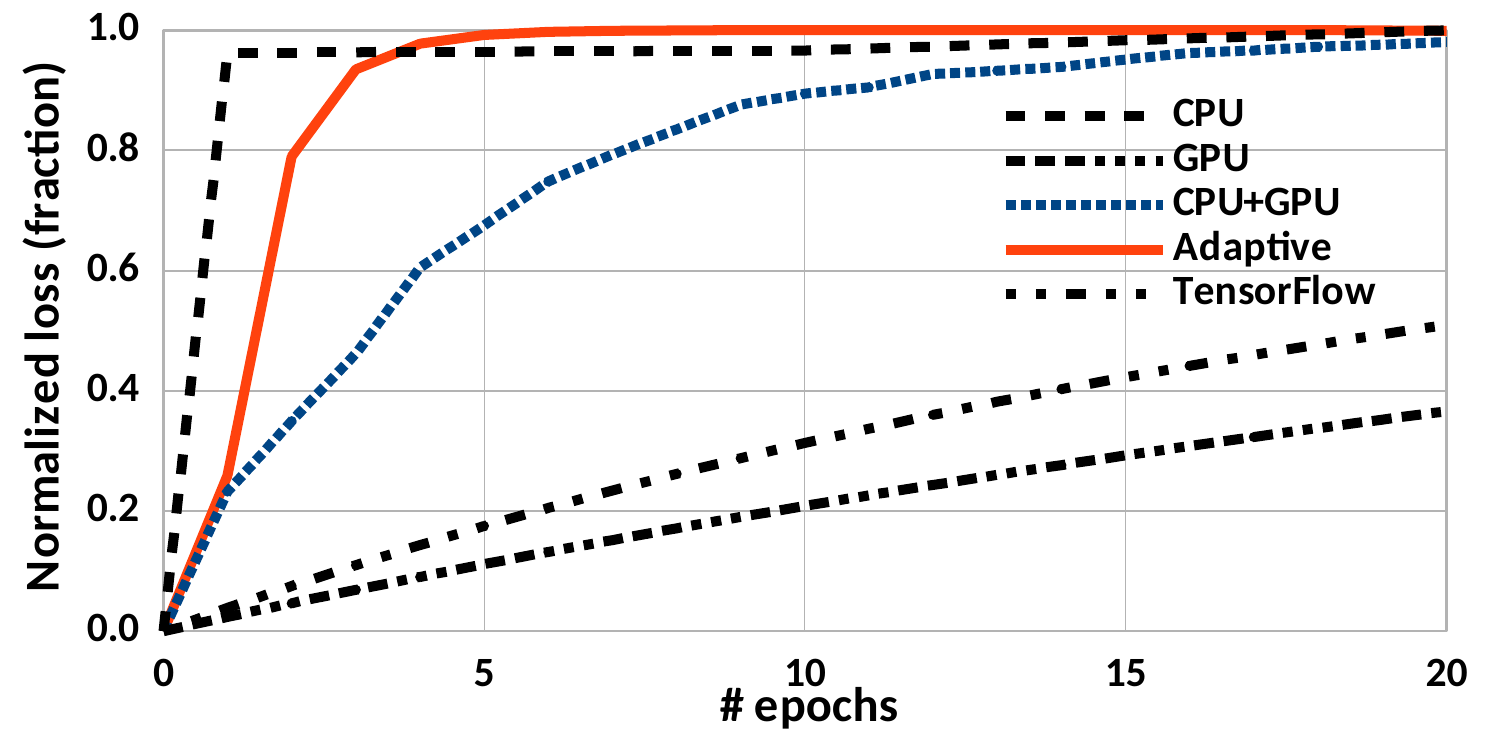}\label{fig:w-iter}}\hfill
\subfloat[w8a V100]{\includegraphics[width=0.47\textwidth]{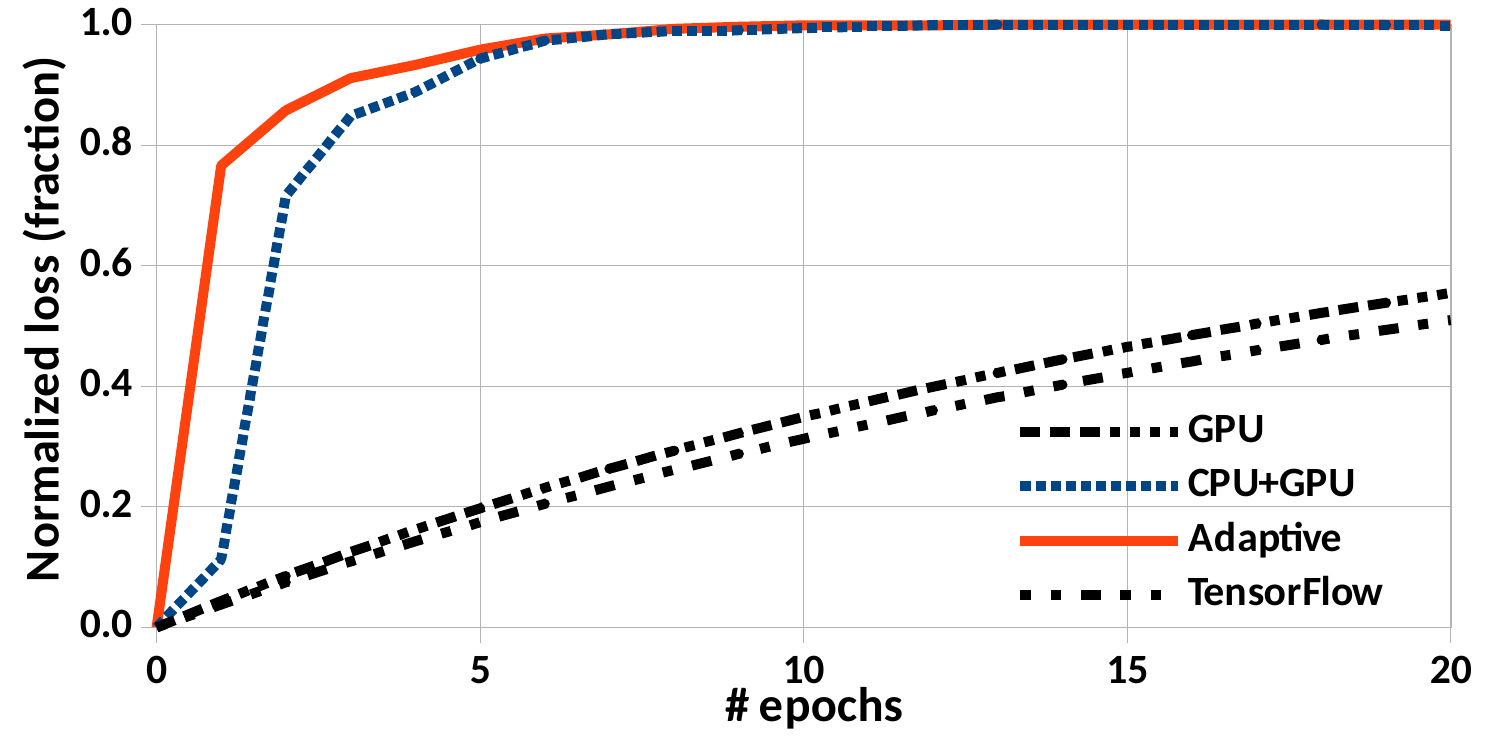}\label{fig:w-iter-aws}} \\
\subfloat[delicious K80]{\includegraphics[width=0.47\textwidth]{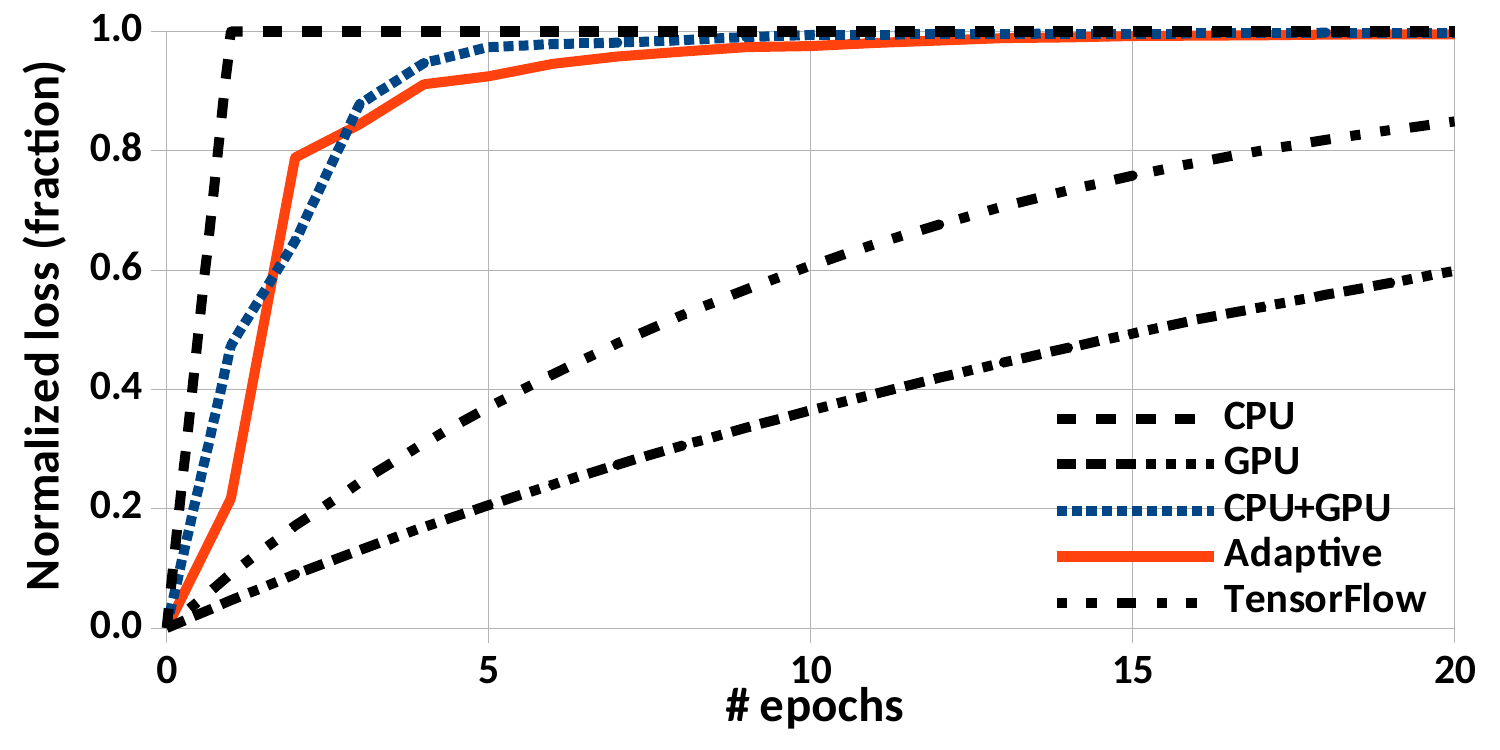}\label{fig:d-iter}}\hfill
\subfloat[delicious V100]{\includegraphics[width=0.47\textwidth]{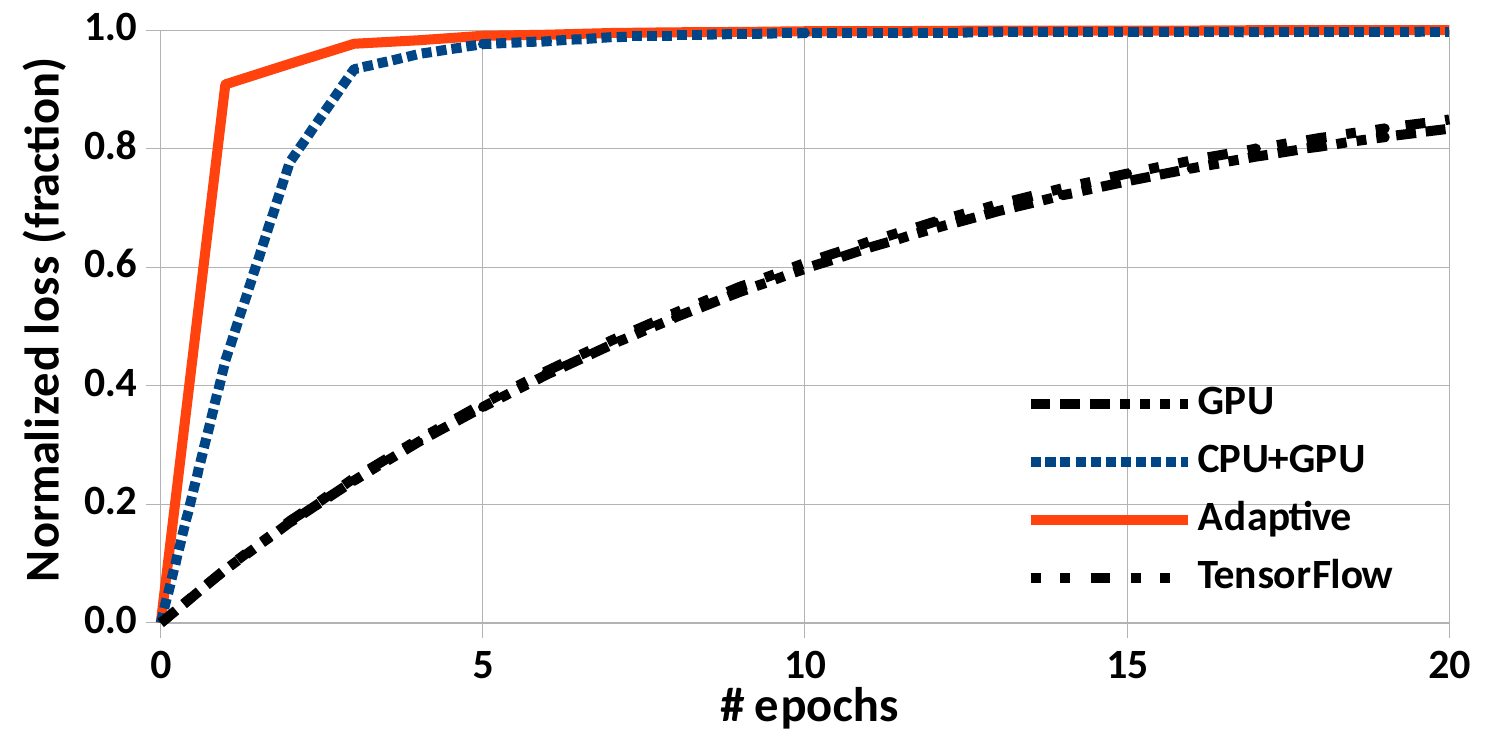}\label{fig:d-iter-aws}} \\
\subfloat[real-sim K80]{\includegraphics[width=0.47\textwidth]{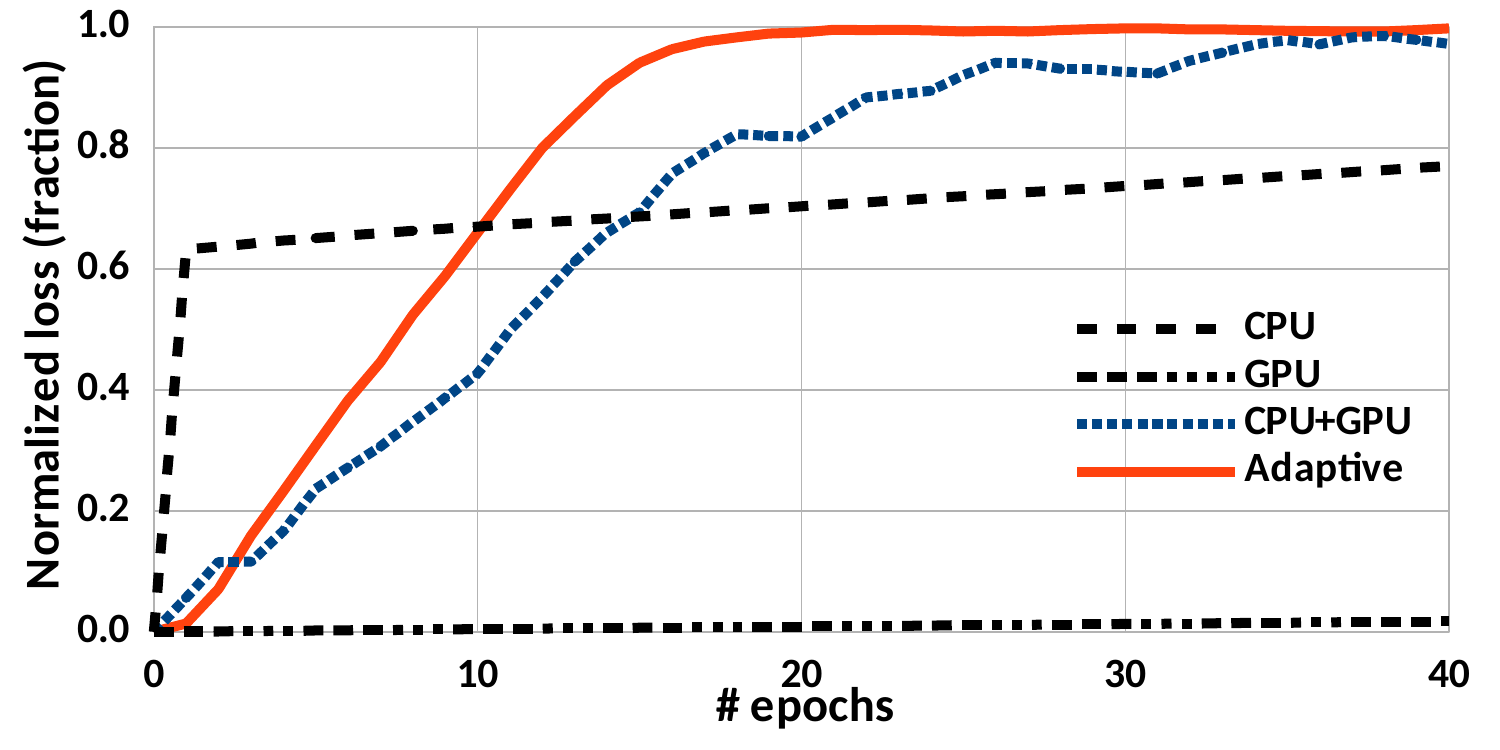}\label{fig:r-iter}}\hfill
\subfloat[real-sim V100]{\includegraphics[width=0.47\textwidth]{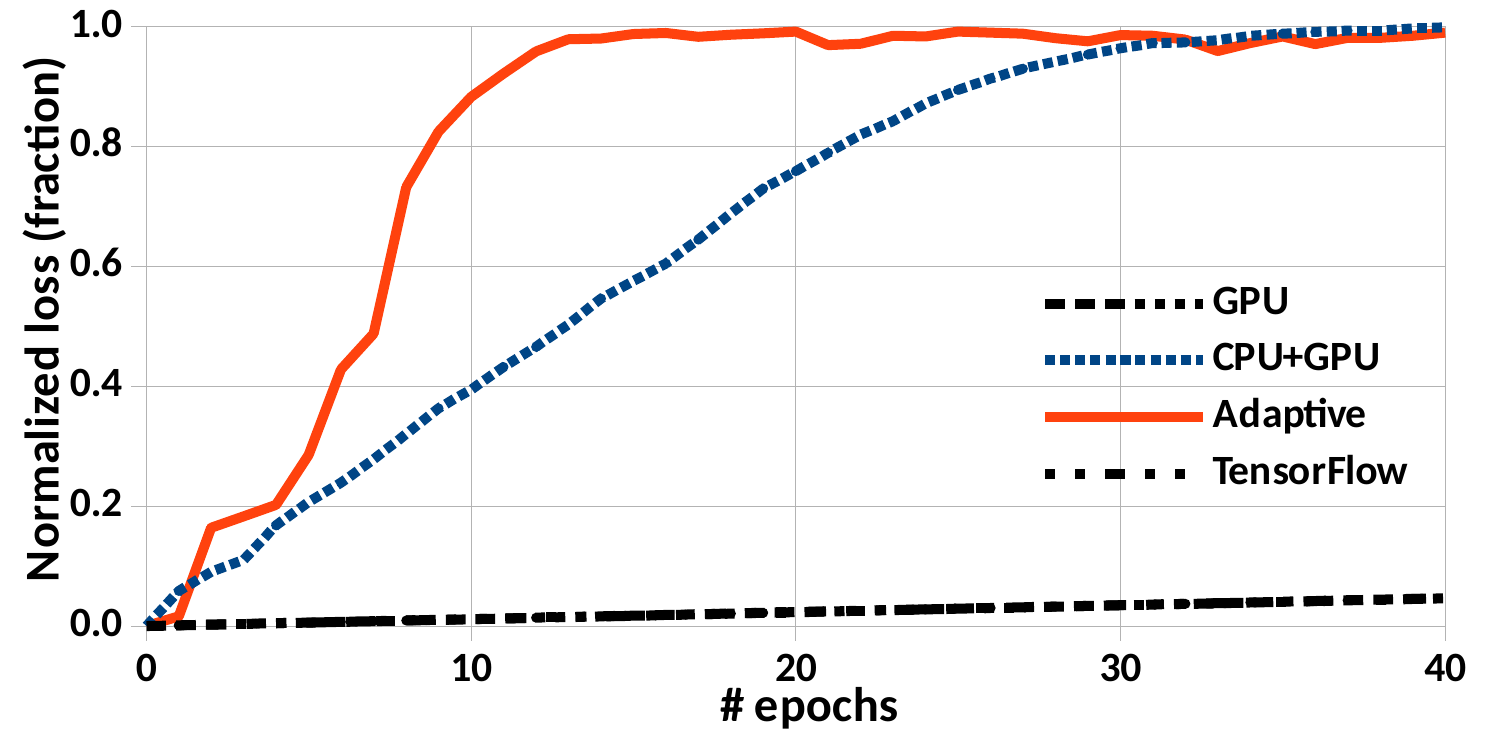}\label{fig:r-iter-aws}}
\caption{Normalized loss for epochs to convergence on the UC Merced server (K80) and the AWS p3.16xlarge instance (V100). Left side is for UC Merced and right side is for AWS.}\label{fig:loss-as-iter}
\end{center}
\end{figure}

\paragraph*{Statistical efficiency}
Figure~\ref{fig:loss-as-iter} depicts the statistical efficiency corresponding to the time to convergence in Figure~\ref{fig:loss-as-time}. Statistical efficiency -- or loss convergence as a function of the number of epochs -- is directly proportional with the number of effective model updates per epoch. The more updates an algorithm performs, the better its statistical efficiency is. Since the number of updates per epoch is given by the number of processed batches, we expect that smaller batches provide better statistical efficiency. This is confirmed by the Hogwild (CPU) results on the UC Merced server which outperforms the other algorithms in almost all the cases. The only exception is the real-sim dataset on which a significant decrease in loss is obtained fast, followed by a much slower linear trajectory to a sub-optimal minimum. The reason we do not include Hogwild (CPU) results on the AWS instance is financial. The cost per hour is upward of \$20 USD, while the time to perform a Hogwild epoch is more than two orders of magnitude higher than for the other algorithms. (Mini-) Hogbatch (GPU) and TensorFlow have the largest batches, thus, they have relatively poor statistical efficiency. The overlapped curves on the AWS instance confirm that our implementation and TensorFlow are identical. This is not the case on the UC Merced server where Hogbatch (GPU) runs on two GPUs, while TensorFlow uses a single GPU. Due to model update conflicts, Hogbatch (GPU) performs slightly worse, especially on the low-dimensional datasets. Since the heterogeneous Hogbatch algorithms combine small and large batches -- as expected -- their efficiency is a weighted average of the two. The larger the gap between the batch sizes, the higher the deviation from the optimal statistical efficiency. This explains the superiority of Adaptive over CPU+GPU. In certain cases -- most notably the covtype and real-sim datasets -- Adaptive achieves similar or better statistical efficiency to Hogwild (CPU).

\paragraph*{Computing architecture impact on convergence}
When we compare the results in Figure~\ref{fig:loss-as-time} across the two servers, we observe a similar trend across algorithms. With a few exceptions, the relative order between the different solutions is maintained. This proves that our implementations are not highly-sensitive to the platform on which they are executed. Across platforms, there is a certain level of variation between CPU and GPU. The CPU implementations on the UC Merced server achieve similar levels of convergence slightly faster, even though they use a slightly smaller number of threads. This is likely due to the fact that we have complete control over this on-premises resource which we utilize in isolation, while the CPUs on the AWS instance are shared. A less likely reason may be the higher number of model update conflicts due to the slightly larger number of threads on the AWS instance. Due to the more powerful V100 GPU, all the algorithms that use the GPU achieve slightly faster convergence on the AWS instance. As opposed from the CPUs, the GPU is not shared in AWS. The important observation is that these differences cancel out for the heterogeneous CPU+GPU algorithms and they achieve the fastest convergence on both architectures.

\begin{figure}[htbp]
\begin{center}
\subfloat[UC Merced]{\includegraphics[width=.5\textwidth]{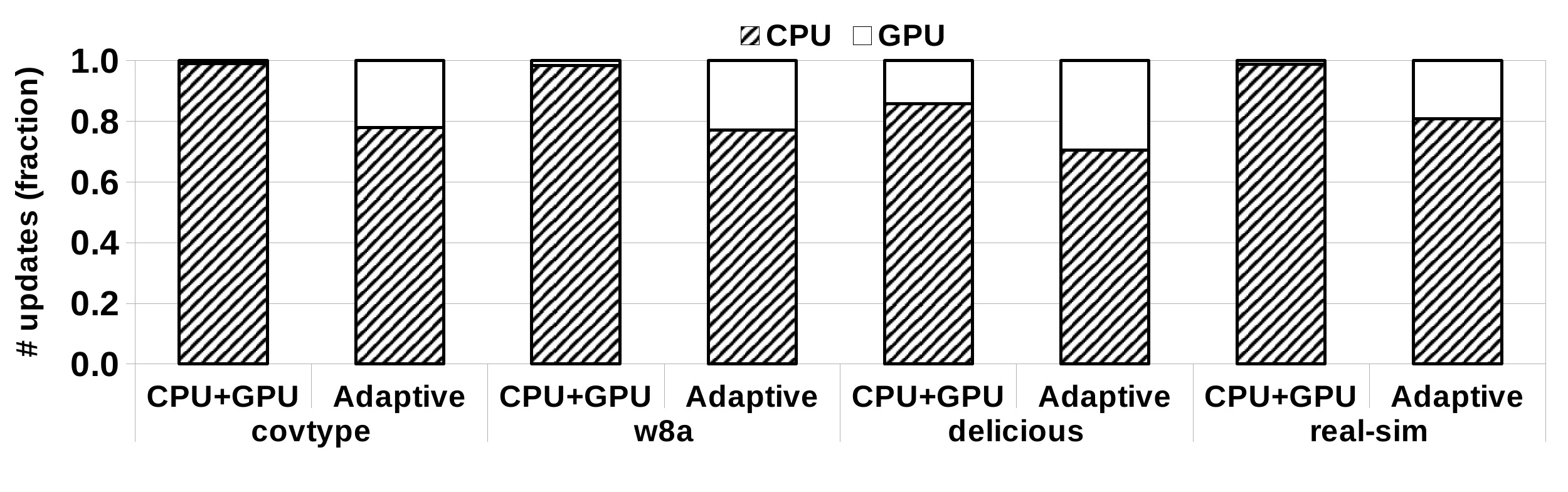}\label{fig:updates-ucm}}\hfill
\subfloat[AWS p3.16xlarge]{\includegraphics[width=.5\textwidth]{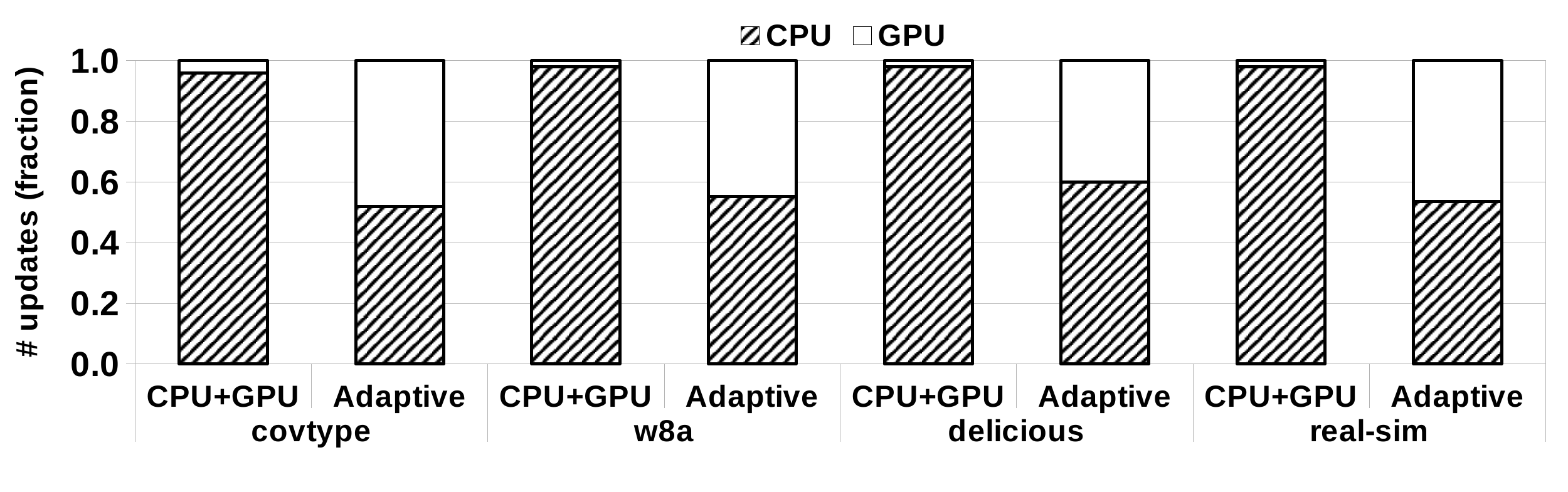}\label{fig:updates-aws}}
\caption{Ratio of model updates applied by CPU and GPU.}\label{fig:updates}
\end{center}
\end{figure}

\paragraph*{Model updates distribution}
The ratio of model updates performed by the CPU and GPU in the heterogeneous Hogbatch algorithms is depicted in Figure~\ref{fig:updates}. In the case of CPU+GPU, the CPU updates are almost exclusive because the gap between the batch size on CPU and GPU is maximized---the CPU batch size is 1, while the GPU batch size is in the order of thousands. As the gap decreases, the distribution moves towards uniformity, with each of the CPU and GPU performing approximately half of the updates in Adaptive on the AWS instance. Since the UC Merced server uses two GPUs, the ratio of updates performed by a single GPU is smaller---the figure shows the updates performed by a single GPU. If we aggregate the updates performed by both GPUs, the results are similar to the AWS case. It is important to notice that the CPU and GPU updates have different weight. The CPU updates are applying coarse gradients computed over small batches---as small as a single training example. The GPU updates are computed over a much larger batch. Thus, the corresponding gradients are more accurate. Since the number of examples in the training dataset is constant and there are 48/56 CPU threads performing model updates compared to a single GPU, for an evenly split model update ratio, the GPU processes 48/56 more batches than the CPU---and these are considerably larger batches. The goal of the Adaptive algorithm is to control the batch size such that this balanced split between the CPU and GPU updates is achieved, independent of the initial CPU and GPU batch sizes. Essentially, Adaptive frees the user from the burden to find an appropriate step size.

\begin{figure*}[htbp]
\centering
\includegraphics[width=\textwidth]{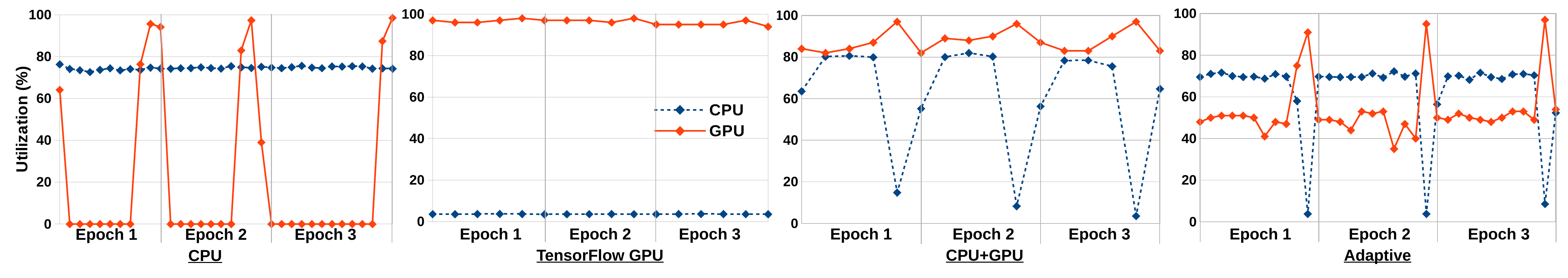}
\caption{CPU and GPU utilization for three epochs of the Hogbatch algorithms executed on the covtype dataset on the UC Merced server.}\label{fig:util}
\end{figure*}

\paragraph*{Resource utilization}
The CPU and GPU utilization for the execution of three epochs of the Hogbatch algorithms on the covtype dataset -- on the UC Merced server -- are depicted in Figure~\ref{fig:util}---the results for the other datasets follow a similar pattern. The loss computation is always performed on the GPU at the end of the epoch. This explains the increase in GPU utilization -- and the decrease in CPU -- across all the algorithms. The CPU utilization hovers around 80\% because only 48 and 56 of the available 56 and 64 threads, respectively, are used. The slight decrease on Adaptive is due to the larger -- and continuously changing -- batch sizes. The GPU utilization is above 80\% in GPU and CPU+GPU since the batch size is 8,192. The batch size in Adaptive decreases to the lower threshold, which triggers the corresponding decrease in utilization. The lower threshold parameter controls the tradeoff between GPU utilization and convergence. In the case of CPU+GPU, utilization is maximized. In Adaptive, the even distribution of model updates across CPU and GPU is more important. Independent of which approach is taken, the ultimate benefit is the faster time to convergence.

\subsection{Summary}\label{sec:experiments:summary}

The following insights can be derived from the experiments. Both heterogeneous Hogbatch algorithms outperform the CPU and GPU-only solutions in time to convergence by large margins. This is also the case for TensorFlow, which is a GPU-only variant. Due to the much larger number of model updates, Hogwild CPU has the best statistical efficiency. Nonetheless, the Adaptive CPU+GPU algorithm comes within similar performance for all the datasets. The heterogeneous algorithms provide consistent performance across two different computing architectures with different number of GPUs and GPU type. The batch size threshold controls the difference between CPU+GPU and Adaptive both in number of model updates and utilization. These have a direct impact on the convergence of the loss function. With few exceptions, for low-dimensional datasets, CPU+GPU is superior, while Adaptive is better for sparse high-dimensional data.

\section{CONCLUSIONS AND FUTURE WORK}\label{sec:conclusions}

In this paper, we introduce a generic deep learning framework that exploits the difference in computational power and memory hierarchy between CPU and GPU. We design two heterogeneous SGD algorithms based on insights gained from experimentation with the framework. The first algorithm -- CPU+GPU Hogbatch -- combines small batches on CPU with large batches on GPU in order to maximize the utilization of both resources. The second algorithm -- Adaptive Hogbatch -- assigns batches with continuously evolving size based on the relative speed of CPU and GPU. We show that the implementation of these algorithms in the proposed CPU+GPU framework consistently achieves both faster convergence and higher resource utilization than TensorFlow on several real datasets and on two computing architectures. In future work, we plan to scale these algorithms to multi-GPU architectures, beyond the dual Tesla K80 available on the UC Merced server. We also plan to investigate if the proposed algorithms extend to sparse datasets.

\paragraph*{Acknowledgments}
This work is supported by a U.S. Department of Energy Early Career Award (DOE Career).

\bibliographystyle{abbrv}

\begin{thebibliography}{11}

\bibitem{amd-cpu}
{AMD EPYC 7742}.
\newblock \url{https://www.amd.com/en/products/cpu/amd-epyc-7742}.
[Accessed March 2020]

\bibitem{aws-ec2-gpu}
{Amazon EC2 Instance Types}.
\newblock \url{https://aws.amazon.com/ec2/instance-types/}.
[Accessed March 2020]

\bibitem{cpu+gpu-code}
{CPU+GPU SGD Code}.
\newblock \url{https://github.com/YMA33/CPU-GPU-SGD}.
[Accessed April 2020]

\bibitem{intel-mkl}
{Intel Math Kernel Library}.
\newblock \url{https://software.intel.com/en-us/mkl}.
[Accessed March 2020]

\bibitem{intel-cpu}
{Intel Xeon Platinum 9282 Processor}.
\newblock
  \url{https://ark.intel.com/content/www/us/en/ark/products/194146/intel-xeon-platinum-9282-processor-77m-\\cache-2-60-ghz.html}.
[Accessed March 2020]

\bibitem{cublas}
{Nvidia cuBLAS}.
\newblock \url{https://developer.nvidia.com/cublas}.
[Accessed March 2020]

\bibitem{openmp}
{OpenMP}.
\newblock \url{https://www.openmp.org/}.
[Accessed March 2020]

\bibitem{perlmutter}
{Perlmutter}.
\newblock \url{https://www.nersc.gov/systems/perlmutter/}.
[Accessed March 2020]

\bibitem{slide-code}
{SLIDE}.
\newblock \url{https://github.com/keroro824/HashingDeepLearning}.
[Accessed March 2020]

\bibitem{summit}
{Summit}.
\newblock
  \url{https://www.olcf.ornl.gov/olcf-resources/compute-systems/summit/}.
[Accessed March 2020]

\bibitem{hpc-top500}
{TOP 500---The List}.
\newblock \url{https://www.top500.org/}.
[Accessed March 2020]

\bibitem{titan}
{Titan}.
\newblock
  \url{https://www.olcf.ornl.gov/olcf-resources/compute-systems/titan/}.
[Accessed March 2020]

\bibitem{tensorflow}
M. Abadi, P.~Barham, J.~Chen, Z.~Chen, A.~Davis, J.~Dean, M.~Devin, S.~Ghemawat, G.~Irving, M.~Isard, M.~Kudlur, J.~Levenberg, R.~Monga, S.~Moore, D.~G. Murray, B.~Steiner, P.~Tucker, V.~Vasudevan, P.~Warden, M.~Wicke, Y.~Yu, and X.~Zheng.
\newblock {TensorFlow: {A} System for Large-Scale Machine Learning}.
\newblock In {\em OSDI 2016}.

\bibitem{bertsekas:igd}
D.~P. Bertsekas.
\newblock {Incremental Gradient, Subgradient, and Proximal Methods for Convex
  Optimization: A Survey}.
\newblock MIT 2010.

\bibitem{Xclassification-dataset-repo}
K.~Bhatia, K.~Dahiya, H.~Jain, A.~Mittal, Y.~Prabhu, and M.~Varma.
\newblock {The Extreme Classification Repository: Multi-label Datasets and
  Code}.
\newblock \url{http://manikvarma.org/downloads/XC/XMLRepository.html}, 2016.
[Accessed March 2020]

\bibitem{deep-nets-sgd}
L.~Bottou.
\newblock {\em {Neural Networks: Tricks of the Trade}}.
\newblock Springer, 2012.

\bibitem{bottou:optim-methods-scale-ml}
L.~Bottou, F.~Curtis, and J.~Nocedal.
\newblock {Optimization Methods for Large-Scale Machine Learning}.
\newblock {\em SIAM Review}, 60(2):223--311, 2018.

\bibitem{sync-vs-asynch-sgd}
J.~Chen, R.~Monga, S.~Bengio, and R.~J{\'{o}}zefowicz.
\newblock {Revisiting Distributed Synchronous {SGD}}.
\newblock {\em CoRR}, abs/1604.00981, 2016.
\newblock \url{http://arxiv.org/abs/1604.00981}

\bibitem{geeps}
H.~Cui, H.~Zhang, G.~R. Ganger, P.~B. Gibbons, and E.~P. Xing.
\newblock {GeePS: Scalable Deep Learning on Distributed GPUs with a
  GPU-specialized Parameter Server}.
\newblock In {\em {EuroSys 2016}}.

\bibitem{buckwild}
C.~{De Sa}, M.~Feldman, K.~Olukotun, and C.~R{\'e}.
\newblock {Understanding and Optimizing Asynchronous Low-Precision Stochastic
  Gradient Descent}.
\newblock In {\em {ISCA 2017}}.

\bibitem{finance-sgd}
M.~Dixon, D.~Klabjan, and J.~Bang.
\newblock {Classification-based Financial Markets Prediction using Deep Neural
  Networks}.
\newblock {\em CoRR}, abs/1603.08604, 2016.
\newblock \url{http://arxiv.org/abs/1603.08604}

\bibitem{facebook:large-batches}
P.~Goyal, L.~Wesolowski, P.~Dollar, A.~Kyrola, R.~Girshick, A.~Tulloch,
  P.~Noordhuis, Y.~Jia, and K.~He.
\newblock {Accurate, Large Minibatch SGD: Training ImageNet in 1 Hour}.
\newblock {\em CoRR}, abs/1706.02677v2, 2018.
\newblock \url{http://arxiv.org/abs/1706.02677}

\bibitem{omnivore}
S.~Hadjis, C.~Zhang, I.~Mitliagkas, and C.~R{\'{e}}.
\newblock {Omnivore: An Optimizer for Multi-device Deep Learning on CPUs and
  GPUs}.
\newblock {\em CoRR}, abs/1606.04487, 2016.
\newblock \url{http://arxiv.org/abs/1606.04487}

\bibitem{cvpr-sgd}
K.~He, X.~Zhang, S.~Ren, and J.~Sun.
\newblock {Deep Residual Learning for Image Recognition}.
\newblock In {\em CVPR 2016}.

\bibitem{speech-sgd}
G.~Hinton, L.~Deng, D.~Yu, G.~Dahl, and A.~Mohamed.
\newblock {Deep Neural Networks for Acoustic Modeling in Speech Recognition}.
\newblock {\em IEEE Signal Processing Magazine}, 29:82--97, 2012.

\bibitem{flex-ps}
Y.~Huang, T.~Jin, Y.~Wu, Z.~Cai, X.~Yan, F.~Yang, J.~Li, Y.~Guo, and J.~Cheng.
\newblock {FlexPS: Flexible Parallelism Control in Parameter Server
  Architecture}.
\newblock {\em PVLDB}, 11(5):566--579, 2018.

\bibitem{hetero-ps}
J.~Jiang, B.~Cui, C.~Zhang, and L.~Yu.
\newblock {Heterogeneity-aware Distributed Parameter Servers}.
\newblock In {\em {SIGMOD 2017}}, pages 463--478.

\bibitem{crossbow}
A.~Koliousis, P.~Watcharapichat, M.~Weidlich, L.~Mai, P.~Costa, and
  P.~Pietzuch.
\newblock {CROSSBOW: Scaling Deep Learning with Small Batch Sizes on Multi-GPU
  Servers}.
\newblock {\em PVLDB}, 12(11):1399--1413, 2019.

\bibitem{parameter-server}
M.~Li, L.~Zhou, Z.~Yang, A.~Li, F.~Xia, D.~G. Andersen, and A.~Smola.
\newblock {Scaling Distributed Machine Learning with the Parameter Server}.
\newblock In {\em OSDI 2014}.

\bibitem{bgd-vs-sgd:ipdps-2019}
Y.~Ma, F.~Rusu, and M.~Torres.
\newblock {Stochastic Gradient Descent on Modern Hardware: Multi-core CPU or
  GPU? Synchronous or Asynchronous?}
\newblock In {\em {IPDPS 2019}}.

\bibitem{bgd-vs-sgd:arxiv-2018}
Y.~Ma, F.~Rusu, and M.~Torres.
\newblock {Stochastic Gradient Descent on Highly-Parallel Architectures}
\newblock {\em CoRR}, abs/1802.08800, 2018.
\newblock \url{http://arxiv.org/abs/1802.08800}

\bibitem{hogwild}
F.~Niu, B.~Recht, C.~R{\'e}, and S.~J. Wright.
\newblock {Hogwild: A Lock-Free Approach to Parallelizing Stochastic Gradient
  Descent}.
\newblock In {\em NIPS 2011}.

\bibitem{hogwild-disk}
C.~Qin, M.~Torres, and F.~Rusu.
\newblock {Scalable Asynchronous Gradient Descent Optimization for Out-of-Core
  Models}.
\newblock {\em PVLDB}, 10(10):986--997, 2017.

\bibitem{combustion-dnn}
T.~Ren, M.~F. Modest, A.~Fateev, G.~Sutton, W.~Zhao, and F.~Rusu.
\newblock {Machine Learning Applied to Retrieval of Temperature and
  Concentration Distributions from Infrared Emission Measurements}.
\newblock {\em Applied Energy}, 252(113448), 2019.

\bibitem{sgd:survey}
S.~Ruder.
\newblock {An Overview of Gradient Descent Optimization Algorithms}.
\newblock {\em CoRR}, abs/1609.04747v2, 2017.
\newblock \url{http://arxiv.org/abs/1609.04747}

\bibitem{hogbatch}
S.~Sallinen, N.~Satish, M.~Smelyanskiy, S.~Sury, and C.~R{\'e}.
\newblock {High Performance Parallel Stochastic Gradient Descent in Shared
  Memory}.
\newblock In {\em IPDPS 2016}.

\bibitem{heterogeneous:ieee-micro}
M.~J. Schulte, M.~Ignatowski, G.~H. Loh, B.~M. Beckmann, W.~C. Brantley,
  S.~Gurumurthi, N.~Jayasena, I.~Paul, S.~K. Reinhardt, and G.~Rodgers.
\newblock {Achieving Exascale Capabilities Through Heterogeneous Computing}.
\newblock {\em IEEE Micro}, 35(4):26--36, 2015.

\bibitem{memory-mgmt:ipdps-2019}
S.~B. Shriram, A.~Garg, and P.~Kulkarni.
\newblock {Dynamic Memory Management for GPU-based Training of Deep Neural
  Networks}.
\newblock In {\em {IPDPS 2019}}.

\bibitem{gpu-direct-storage}
A.~Thompson and C.~Newburn.
\newblock {GPUDirect Storage: A Direct Path Between Storage and GPU Memory}.
\newblock \url{https://devblogs.nvidia.com/gpudirect-storage/}, 2019.
[Accessed March 2020]

\bibitem{imagenet-minutes:iclr-2020}
Y.~You, J.~Li, S.~Reddi, J.~Hseu, S.~Kumar, S.~Bhojanapalli, X.~Song,
  J.~Demmel, K.~Keutzer, and C.-J. Hsieh.
\newblock {Large Batch Optimization for Deep Learning: Training BERT in 76
  Minutes}.
\newblock In {\em {ICLR 2020}}.

\bibitem{imagenet-minutes:icpp-2018}
Y.~You, Z.~Zhang, C.-J. Hsieh, J.~Demmel, and K.~Keutzer.
\newblock {ImageNet Training in Minutes}.
\newblock In {\em {ICPP 2018}}.

\bibitem{dimm-witted}
C.~Zhang and C.~R{\'e}.
\newblock {DimmWitted: A Study of Main-Memory Statistical Analytics}.
\newblock {\em PVLDB}, 7(12), 2014.

\bibitem{asynch-sgd-delay-comp}
S.~Zheng, Q.~Meng, T.~Wang, W.~Chen, N.~Yu, Z.~Ma, and T-Y.~Liu.
\newblock {Asynchronous Stochastic Gradient Descent with Delay Compensation for
  Distributed Deep Learning}.
\newblock {\em CoRR}, abs/1609.08326, 2016.
\newblock \url{http://arxiv.org/abs/1609.08326}

\end{thebibliography}

\end{document}